\documentclass[12pt]{iopart}
\usepackage{iopams}  

\begin{document}

\title{Backreaction from non-conformal quantum fields in de Sitter spacetime}

\author{Guillem P\'erez-Nadal$^1$, Albert Roura$^2$ and Enric Verdaguer$^1$}

\address{$^1$Departament de F\'{\i}sica Fonamental
and Institut de Ci\`encies del Cosmos, Universitat de Barcelona,
Av.~Diagonal 647, 08028 Barcelona, Spain}
\address{$^2$Theoretical Division, T-8, Los Alamos National Laboratory,
M.S.~B285, Los Alamos, NM 87545, USA}

\begin{abstract}
We study the backreaction on the mean field geometry due to a non-conformal quantum field in a Robertson-Walker background. In the regime of small mass and small deviation from conformal coupling, we compute perturbatively the expectation value of the stress tensor of the field for a variety of vacuum states, and use it to obtain explicitly the semiclassical gravity solutions for isotropic perturbations around de Sitter
spacetime, which is found to be stable.
Our results show clearly the crucial role of the non-local terms that appear in the effective action: they cancel the contribution from local terms proportional to the logarithm of the scale factor which would otherwise become dominant at late times and prevent the existence of a stable self-consistent de Sitter solution.
Finally, the opposite regime of a strongly non-conformal field with a large mass is also considered.
\end{abstract}



\section{Introduction}
\label{sec1}

Our present understanding of cosmology assumes that the universe
underwent a short period of accelerated expansion known as inflation
\cite{guth81,linde82a,albrecht82,linde83a,linde90}.  The inflationary
scenario has been remarkably successful in explaining the observed
anisotropies of the cosmic microwave background
\cite{smoot92,bennett03,peiris03,spergel07}. In most inflationary
models the accelerated expansion phase is close to but never exactly
de Sitter and this phase eventually ends when the kinetic energy of
the inflaton field driving inflation starts to dominate over the
potential term. On the other hand, observations of distant supernovae
indicate that the universe is presently undergoing a period of
accelerated expansion \cite{perlmutter99,riess98} that may be driven
by a small non-vanishing cosmological constant
\cite{seljak06,tegmark06,giannantonio06,eisenstein05}. If that is the
case, the geometry of our universe would tend to that of de Sitter
spacetime at sufficiently late times. Thus, a detailed knowledge of
the physics associated with de Sitter space may play a key role in
understanding both the very early universe as well as its ultimate
fate. Furthermore, it is conceivable that studying a possible
screening of the cosmological constant driving de Sitter space, due to
quantum effects, could shed some light on the huge fine-tuning problem
that the current value of the cosmological constant seems to pose.

An open question which has recently received increasing attention is
whether the quantum fluctuations of the metric and the matter fields
in de Sitter space can give rise to large backreaction effects on the
mean background geometry. It has been argued that in pure gravity with
a cosmological constant the infrared effects due to two graviton loops
and higher-order radiative corrections could lead to a secular
screening of the cosmological constant
\cite{tsamis96a,tsamis97}. There have also been proposals that a
significant screening of the cosmological constant could appear in
chaotic inflationary models at one loop when both the metric and
inflaton field fluctuations are considered
\cite{mukhanov97,abramo97,abramo99,losic05}. In all these cases the
quantum fluctuations of the metric play an essential role. However,
whenever the metric perturbations are quantized, one needs to confront
the problem of defining proper diffeomorphism-invariant observables in
quantum gravity \cite{giddings05}, even when treated as a low-energy
effective field theory. In particular one needs to make sure that the
secular screening found in the analysis mentioned above is not simply
a gauge artifact. As a matter of fact, it was shown in
Refs.~\cite{abramo02b,geshnizjani02} that when a suitable
gauge-invariant measure of the expansion rate was considered, the
screening effect previously found in chaotic inflationary models was
not actually present (at least for single field models). Similarly, a
recent reanalysis of the pure gravity case which made use of a
diffeomorphism-invariant measure of the change of the expansion rate
indicated the absence of secular effects to all orders in perturbation
theory \cite{garriga08} (although this conclusion is still subject to
certain debate \cite{tsamis07}).

In fact, the study of backreaction effects from quantum matter fields in de Sitter spacetime has a long history. Fischetti, Hartle and Hu employed effective action methods to study the quantum backreaction of conformal fields on the dynamics of a Robertson-Walker (RW) spacetime \cite{fischetti79}. Shortly afterwards Starobinsky showed that the vacuum polarization effects of a large number of conformal fields could drive a de Sitter expansion stage without the need for a classical cosmological constant and that this de Sitter solution was unstable under RW-type perturbations (for a certain sign of one of the parameters in the effective action), which provided a ``graceful exit'' mechanism towards a standard cosmological evolution \cite{starobinsky80,vilenkin85}. This scenario is often known as \emph{Starobinsky} (or \emph{trace anomaly}) \emph{inflation}. However, it has later been argued that within an effective field theory (EFT) approach (which provides a natural framework for understanding this kind of calculations) the higher-order curvature terms in the effective action should be treated perturbatively
\cite{simon91,parker93,flanagan96,burgess04}. When doing so, the de Sitter solution entirely driven by the vacuum polarization of the conformal fields is no longer present \cite{simon92},%
\footnote{One could argue that since the Hubble radius in Starobinsky inflation is of order $\sqrt{N} l_\mathrm{P}$ (where $N$ is the number of conformal fields), for a sufficiently large $N$ the quantum gravity corrections should be small. Nevertheless, the same kind of argument would imply the existence of instabilities for perturbations around flat space with a characteristic time-scale of order $\sqrt{N} l_\mathrm{P}$ \cite{flanagan96}. Moreover, in particular implementations of Starobinsky inflation where the AdS/CFT correspondence can be applied \cite{hawking01}, the scale where higher order corrections become relevant and the low-energy effective field theory expansion breaks down is actually $\sqrt{N} l_\mathrm{P}$ rather than $l_\mathrm{P}$.}
and a de Sitter solution driven by a cosmological constant $\Lambda$ is stable even when the vacuum polarization effects from the conformal fields are included.
More recently, it has been argued that the backreaction from non-conformal fields can lead to significant deviations from a de Sitter solution (in some cases this has been analyzed in the context of trace anomaly inflation \cite{shapiro02,pelinson03a}, but it would also apply to a $\Lambda$-driven de Sitter spacetime). Similarly, it has also been concluded that graviton one-loop effects can lead to significant deviations from de Sitter for the background dynamics \cite{espriu05,cabrer07}. Since the metric perturbations are quantized in this case, there are gauge ambiguities, as mentioned above, which need to be understood and properly addressed. However, if one temporarily ignores this point, the resulting effective action which governs the background dynamics has the same form as that describing the vacuum polarization effects of non-conformal quantum matter fields.

Here we reanalyze this problem by explicitly solving the backreaction on the mean gravitational field due to the quantum effects of
a non-conformal scalar field when the quantum
fluctuations of the metric are not considered. This kind of one-loop
calculation is entirely equivalent to studying the corresponding
backreaction problem in the semiclassical gravity framework
\cite{birrell94,wald94,flanagan96} by solving the
semiclassical Einstein equation, which includes the suitably
renormalized quantum expectation value of the stress tensor operator
acting as a source.  Specifically, in our calculation we assume the
presence of a cosmological constant, which would lead to a de Sitter
solution in the absence of quantum effects, and simplify the problem
by focusing on RW geometries, corresponding to spatially homogenous
and isotropic states of the quantum field.

There exist relevant antecedents to our analysis in the context of both
quantum field theory in a fixed curved spacetime (when
the backreaction of the quantum fields on the spacetime geometry is
not taken into account) and semiclassical gravity. The so-called Bunch-Davies vacuum
\cite{bunch78a,birrell94} for fields in de Sitter is a state invariant
under all the isometries of de Sitter space, which is maximally
symmetric. The renormalized expectation value of the stress tensor
operator for that state is proportional (with a constant factor) to
the metric and, therefore, its contribution to the semiclassical
Einstein equation has the same form as a cosmological constant
term, which allows the existence of self-consistent semiclassical
de Sitter solutions \cite{dowker76,starobinsky80,wada83}.
More importantly, it was shown in Ref.~\cite{anderson00} that
for fields with a wide range of mass and curvature-coupling parameters
evolving in a given de Sitter spacetime, the expectation value of the
stress tensor for any reasonable initial state
tended at late times to the expectation value for the Bunch-Davies
vacuum, where by reasonable states one means states with the same
ultraviolet behavior as the Minkowski vacuum, i.e. with
essentially no excitations at arbitrarily high frequencies (technically they are known as fourth-order adiabatic states \cite{birrell94,habib00}). This
result can be intuitively understood as follows: the exponential
expansion will redshift any finite frequency excitations of the
Bunch-Davies vacuum so that their contribution to the stress tensor
will tend to zero at late times.
This result suggests that even when
taking into account backreaction effects, perturbations around de
Sitter will be redshifted away and at late times the spacetime
geometry will approach de Sitter space with an effective cosmological
constant which includes the contribution from the expectation value of
the stress tensor for the Bunch-Davies vacuum. Such an expectation is supported by the analysis of \cite{isaacson91,rogers92,busch92}, where the linearized semiclassical Einstein equation for spatially isotropic perturbations was considered. The asymptotic behavior of its solutions at late times was analyzed and found to be stable.

Our approach enables us to obtain explicit and relatively simple results for the solutions of the semiclassical backreaction equation at all times, rather than just study its asymptotic stability. This can be achieved for the two opposite regimes of \emph{weakly} and \emph{strongly} non-conformal fields. The weakly non-conformal case corresponds to fields with a mass much smaller than the Hubble parameter and a small deviation from conformal coupling to the curvature. In that case one can treat perturbatively the mass and the parameter characterizing the deviation from conformal coupling. Our results for the semiclassical solutions, given by (\ref{3.30}) and (\ref{3.42}), confirm the stability of de Sitter for weakly non-conformal fields.
This extends our results in an earlier study restricted to massless fields with a slightly non-conformal coupling to the curvature~\cite{perez-nadal07}, where a similar method was employed.
On the other hand, for the strongly non-conformal case, corresponding to a Compton wavelength much smaller than the typical curvature radius, one can introduce a quasi-local approximation that gives rise to a local expansion for the effective action involving positive powers of the curvature divided by the mass squared. We find that through order $1/m^2$ no non-trivial spatially isotropic perturbations are allowed.
These approximation schemes do not provide information on the intermediate regime of masses comparable to the Hubble parameter, but make it possible to obtain explicit expressions with a simple form for the other two regimes. One of the advantages of our explicit results is that they clearly show the crucial role played by the non-local terms (particularly for the weakly non-conformal case). Indeed, the effective action and the backreaction equation that one can derive from it exhibit local terms proportional to the logarithm of the scale factor which become dominant at late times and can cause substantial effects. However, these terms are canceled out by a similar contribution from the non-local terms. This is an important point which seems to have been unnoticed in previous studies and can have significant implications on the validity of local approximations where the non-local terms are neglected.

The plan of the paper is the following. In section~\ref{sec2} we introduce the model and the class of  initial states that we will be considering. We also explain how to obtain the expectation value of the stress tensor from the influence action and how to renormalize the divergences that appear in the influence action. Finally, we derive the backreaction equation and its solutions in terms of the expectation value of the stress tensor. The effective action for weakly non-conformal fields is computed perturbatively in section~\ref{sec3} and explicit results for the stress tensor and the solutions of the backreaction equation are obtained for massless non-conformally coupled fields and massive conformally coupled ones. A more detailed exposition of certain technical aspects can be found in \cite{perez-nadal07}, where the massless non-conformally coupled case was studied. In section~\ref{sec4} we describe how to obtain the influence action for strongly non-conformal fields with a large mass using a quasi-local expansion, and their backreaction effects on small perturbations around de Sitter. We also explain the relevance of these results in order to compare with certain dark energy models based on the renormalization group (RG) running of the cosmological constant.
Finally, we discuss the implications of our results in section~\ref{sec5}.
The relationship between the class of states that we consider and the fourth-order adiabatic vacua for RW spacetimes is explained in the appendix.
Throughout the paper we use natural units with $\hbar=c=1$ and the
$(+,+,+)$ convention of~\cite{misner73}.

\section{General model}
\label{sec2}

We consider a spatially flat RW metric, which in conformal time has the form
\begin{equation}
g_{\mu\nu}=a^2(\eta)\eta_{\mu\nu},
\label{2.1}
\end{equation}
where $\eta_{\mu\nu}={\textrm {diag}}(-1,1,\dots,1)$ is the $n$-dimensional Minkowski metric (we use arbitrary dimensions for the moment in order to perform dimensional regularization later on), and $a(\eta)$ is the scale factor in terms of the conformal time $\eta$. We will study the dynamics of a free quantum scalar field of mass $m$ in such a spacetime, whose action is
\begin{equation}
S[a,\Phi]=-\frac{1}{2}\int d^n x\, a^n\left\{a^{-2}\,\eta^{\mu\nu}\partial_\mu\Phi\partial_\nu\Phi+[m^2+(\xi_c+\nu) R]\Phi^2\right\},
\label{2.2}
\end{equation}
where the dimensionless parameters $\xi_c=(n-2)/[4(n-1)]$ (equal to 1/6 in four dimensions) and $\nu$ give the coupling to $R$, which is the Ricci curvature scalar associated with the metric (\ref{2.1}). It is given by
\begin{equation}
R=2(n-1)\left(\frac{\ddot a}{a^3}+\frac{n-4}{2}\frac{\dot a^2}{a^4}\right).
\label{2.2b}
\end{equation}
Here and throughout the rest of the paper overdots denote
derivatives with respect to the conformal time, i.e. $\dot{}
\equiv d/d\eta$. The case $\nu=0$ is the so-called conformal coupling. If the field is massless and conformally coupled, then it is said to be conformally invariant, or simply conformal. In this paper we will be interested in the general case of non-conformal fields. It is convenient to introduce the rescaled scalar field $\phi(x)=a^{(n-2)/2}(\eta)\Phi(x)$. In terms of this rescaled field, the action simplifies to 
\begin{equation}
S[a,\phi]=-\frac{1}{2}\int d^n x
\left(\eta^{\mu\nu}\partial_\mu\phi\partial_\nu\phi+M^2\phi^2\right),
\label{2.3}
\end{equation}
where $M^2$ is a function of the conformal time given by $M^2(\eta)=a^2(\eta)[m^2+\nu R(\eta)]$. Setting the variation of this action with respect to $\phi$ equal to zero yields the dynamical equation for the rescaled field,
\begin{equation}
(\eta^{\mu\nu}\partial_\mu\partial_\nu-M^2)\phi=0,
\label{2.4}
\end{equation}
which is the usual Klein-Gordon equation in Minkowski spacetime with a time-dependent mass term which vanishes if the field is conformal.

\subsection{The state of the quantum field}
\label{sec2.1}

Let us discuss the kind of states that we will consider for the scalar field $\phi$. The Klein-Gordon equation (\ref{2.4}) admits the following complete set of orthonormal solutions:
\begin{equation}
u_{\vec{k}}(\eta,\vec x)=f_k(\eta)\frac{1}{(2\pi)^{(n-1)/2}}e^{i\vec k\cdot\vec x},
\label{2.5}
\end{equation}
with $\vec k \in \mathbb{R}^{n-1}$ and $k=|\vec k|$. The mode functions $f_k$ satisfy the equation
\begin{equation}
\ddot f_k+(k^2+M^2)f_k=0.
\label{2.6}
\end{equation}
In order for the set $\{u_{\vec{k}}\}$ to be orthonormal, the mode functions must also satisfy the Wronskian condition
\begin{equation}
\dot f_kf_{k}^*-f_k\dot f_{k}^*=-i.
\label{2.6b}
\end{equation}
The solution of equation (\ref{2.6}) is unique once we specify initial conditions at some initial time $\eta_{\rm i}$, $f_k(\eta_{\rm i})$ and $\dot f_k(\eta_{\rm i})$, which must also be consistent with (\ref{2.6b}).
Given a set of orthonormal modes, the field
operator can be expanded in terms of the associated creation and
annihilation operators as $\hat{\phi}=\sum_{\vec{k}}
(\hat{a}_{\vec{k}} u_{\vec{k}} + \hat{a}^\dagger_{\vec{k}}
u^*_{\vec{k}})$, and a Fock space based on the vacuum defined by
$\hat{a}_{\vec{k}} |0\rangle = 0$ can be constructed.
Therefore, since different choices of initial conditions for $f_k$ give rise to different sets of mode functions, they define different (homogeneous and isotropic) vacua.
In this paper we will proceed as follows. Define an auxiliary scale factor $a_\Psi(\eta)$, with domain $(-\infty,\eta_{\rm i}]$, such that
\begin{equation}
\lim_{\eta\to-\infty}M^{2}_\Psi=0.
\label{2.7}
\end{equation}
Then the initial conditions for the mode functions will be chosen as
\begin{eqnarray}
f_k(\eta_{\rm i})&=&f_{k}^\Psi(\eta_{\rm i})\nonumber\\
\dot f_k(\eta_{\rm i})&=&\dot f_{k}^\Psi(\eta_{\rm i}),
\label{2.8}
\end{eqnarray}
where $f_{k}^\Psi(\eta)$ is the solution of the mode equation (\ref{2.6}) with $M^{2}_\Psi(\eta)$ replacing $M^{2}(\eta)$,
\begin{equation}
\ddot f_{k}^\Psi+(k^2+M^{2}_\Psi)f_{k}^\Psi=0,
\label{2.9}
\end{equation}
that behaves as a standard plane wave at past infinity, i.e.
\begin{equation}
f_{k}^\Psi(\eta)\to\frac{1}{\sqrt{2k}}e^{-i k\eta}
\label{2.10}
\end{equation}
when $\eta\to-\infty$. The state $|\Psi\rangle$ that we will consider is the vacuum associated with the initial conditions (\ref{2.8}). 
By taking different auxiliary scale factors $a_\Psi$, one can get different initial conditions, and thus different vacua. So far, $a_\Psi$ is completely arbitrary, except for condition (\ref{2.7}). However, in order for the state to have the right ultraviolet behavior, further conditions must be imposed on $a_\Psi$, as we will see below.

The states that we have defined have a remarkable computational advantage: they are just the \emph{in}-vacuum when the scale factor is
\begin{equation}
\bar a(\eta)=\left\{\begin{array}{ll}
a(\eta) & \textrm{for $\eta>\eta_{\rm i}$}\\
a_\Psi(\eta) & \textrm{for $\eta\le\eta_{\rm i}$}.
\end{array}\right.
\label{2.11}
\end{equation}
Therefore, instead of working with the scale factor $a(\eta)$ and the state $|\Psi\rangle$, one can perform the computations regarding the scalar field as if it were in a RW spacetime with scale factor $\bar a(\eta)$ and its state were the \emph{in}-vacuum evolving from $\eta=-\infty$.

\subsection{The expectation value of the stress-energy tensor}
\label{sec2.2}

The trace of the stress tensor operator is obtained by functional differentiation of the classical action (\ref{2.3}) with respect to the scale factor:
\begin{equation}
\frac{\delta S}{\delta a}=a^{n-1}\int d^{n-1}x\,T^{\mu}_\mu.
\label{2.12}
\end{equation} 
Similarly, it is well-known \cite{martin99b,hu04a} that its expectation value in the state $|\Psi\rangle$ can be obtained by functionally differentiating the Feynman-Vernon influence action $S_{\rm IF}^\Psi[a^+,a^-]$, which is defined as
\begin{equation}
e^{iS_{\rm{IF}}^\Psi[a^+,a^-]}= \int {\cal D}\phi^+ {\cal D}\phi^-\rho_\Psi[\phi^+(\eta_{\rmi}),\phi^-(\eta_{\rm i})]
e^{i\left(S[a^+,\phi^+]- S[a^-,\phi^-]\right)},
\label{2.13}
\end{equation}
where $\rho_\Psi$ is the density matrix corresponding to the state $|\Psi\rangle$, and the field con\-fig\-u\-ra\-tions $\phi^+$ and $\phi^-$ are supposed to coincide at some final time $\eta_{\rm f}$. Indeed, functionally differentiating this influence action yields the expectation value of the trace of the stress tensor in the state $|\Psi\rangle$,
\begin {equation}
\left.\frac{\delta S_{\rm IF}^\Psi}{{\delta a}^+}\right|_{a^\pm=a}= a^{n-1}{\cal V}\langle T^\mu_\mu\rangle_\Psi ,
\label{2.17}
\end{equation}
where ${\cal V}$ is the volume of space, ${\cal V}=\int d^{n-1} x$, and we have taken into account that the state $|\Psi\rangle$ is homogeneous. 
Following the discussion around equation (\ref{2.11}), instead of $S_{\rm IF}^\Psi[a^+,a^-]$ we can work with $S_{\rm IF}[\bar a^+,\bar a^-]$, where the absence of the superscript $\Psi$ indicates that the state under consideration is the \emph{in}-vacuum. The latter influence action has the simpler form
\begin{equation}
e^{iS_{\rm{IF}}[\bar a^+,\bar a^-]}= \int {\cal D}\phi^+ {\cal D}\phi^-
e^{i\left(S[\bar a^+,\phi^+]- S^*[\bar a^-,\phi^-]\right)}.
\label{2.18}
\end{equation}
The complex conjugate of the classical action appears because the usual $-i\epsilon$ prescription is used. Since the action (\ref{2.3}) is quadratic in the field, this path integral is Gaussian, and a formal expression for the influence action is readily obtained:
\begin{equation}
S_{\rm{IF}}[\bar a^+,\bar a^-]=-\frac{i}{2}\textrm{tr} \ln G,
\label{2.18b}
\end{equation}
where $G$ is the inverse of the matrix $A={\textrm{diag}}(A_{+}, A_{-})$, with $ A_{+}=\eta^{\mu\nu}\partial_\mu\partial_\nu-\bar M^{2}_+
+i\epsilon$, and $A_{-}=-\left(\eta^{\mu\nu}\partial_\mu\partial_\nu-\bar M^{2}_- -i\epsilon\right)$. The function $ M^{2}(\eta)$ has been defined below equation~(\ref{2.3}). The bar and the subscript $\pm$ just indicate that it corresponds to the scale factor $\bar a^\pm$. Functional differentiation gives
\begin {equation}
\left.\frac{\delta S_{\rm IF}}{{\delta \bar a}^+}\right|_{\bar a^\pm=\bar a}= \bar a^{n-1}{\cal V}\langle \bar T^\mu_\mu\rangle,
\label{2.19}
\end{equation}
where $\langle \bar T^\mu_\mu\rangle$ is the expectation value, in the \emph{in}-vacuum, of the trace of the stress tensor with the scale factor $\bar a(\eta)$. For $\eta>\eta_{\rm i}$, we have
\begin{equation}
\langle \bar T^\mu_\mu(\eta)\rangle=\langle T^\mu_\mu(\eta)\rangle_\Psi .
\label{2.20}
\end{equation}
Since the state $|\Psi\rangle$ is homogeneous and isotropic, the only independent components of the expectation value of the stress tensor are the trace and the $00$ component (the energy density).%
\footnote{Strictly speaking the energy density physically measured by a comoving observer corresponds to $T_{tt}$, which is related to $T_{00}$ by $T_{tt} = a^{-2} T_{00}$. However, from now on we will loosely refer to $T_{00}$ as \emph{energy density} as well.}
In fact, the latter can be obtained from the former. Indeed, using Gauss's theorem together with the equality $\nabla_\mu (\langle \hat{T}^{\mu\nu}\rangle\xi_\nu) = (\dot{a}/a) \langle \hat{T}^\mu_\mu \rangle$ for $\vec\xi=\partial/\partial\eta$, which is a consequence of the stress tensor conservation law and the fact that $\vec\xi$ is a conformal Killing vector field of the RW spacetime, we get
\begin{equation}
\bar a^{n-2}(\eta)\langle \bar T_{00}(\eta)\rangle =-\int^{\eta}_{-\infty}d\eta^\prime  \bar a^{n-1}(\eta^\prime)\dot {\bar a}(\eta^\prime) \langle \bar T^\mu_\mu(\eta^\prime)\rangle,
\label{2.21}
\end{equation}
where we took into account that for a field in the \emph{in}-vacuum $\bar a^{n-2}(\eta)\langle \bar T_{00}(\eta)\rangle$ vanishes in the limit $\eta \to -\infty$.
Again, for $\eta>\eta_{\rm i}$ we have
\begin{equation}
\langle \bar T_{00}(\eta)\rangle=\langle T_{00}(\eta)\rangle_\Psi .
\label{2.22}
\end{equation}

\subsection{Renormalization}
\label{sec2.3}

The influence action (\ref{2.18b}) suffers from ultraviolet divergences in the limit $n\to 4$. The divergent contributions to the influence action can be written as $S_{\rm div}[g^+]-S_{\rm div}[g^-]$, where $S_{\rm div}[g]$ corresponds to the divergent part of the \emph{in-out} effective action. The short-distance behavior of the \emph{in-out} effective action of a free scalar field has been thoroughly studied for a general metric $g$ (not necessarily of RW type) and it has been shown that the divergent part has the following form \cite{birrell94}:
\begin{eqnarray}
S_{\rm div}[g] = -\frac{\mu^{n-4}}{32\pi^2(n-4)}\int d^nx \sqrt{-g}
&& \bigg[m^4+2m^2\nu R+\nu^2R^2 \nonumber \\
&& \, +\frac{1}{90}\left(R_{\mu\nu\alpha\beta} R^{\mu\nu\alpha\beta}
- R_{\mu\nu} R^{\mu\nu}\right)\bigg],
\label{r1}
\end{eqnarray}
where $\mu$ is an arbitrary mass scale introduced for dimensional consistency which plays the role of the renormalization scale when using dimensional regularization; $R_{\mu\nu\alpha\beta}$ and $R_{\mu\nu}$ are the Riemann and the Ricci tensors respectively. The dynamics of the metric in semiclassical gravity can be derived from the following closed-time-path (CTP) effective action \cite{martin99b}:
\begin{equation}
\Gamma_\mathrm{CTP} [g^+,g^-] = S_{\rm g}[g^+] - S_{\rm g}[g^-]
+ S_{\rm IF}[g^+,g^-] ,
\label{ea1}
\end{equation}
where $S_{\rm g}[g]$ is the classical gravitational action, which will be fully specified below. Physical predictions can still be finite provided that the so-called bare parameters of the gravitational action have the appropriate divergent behavior so that they cancel the divergences of the bare influence action $S_{\rm IF}[g^+,g^-]$ and the total effective action $\Gamma_\mathrm{CTP} [g^+,g^-]$ is finite. This is accomplished by considering a bare gravitational action of the form
\begin{equation}
S_{\rm g}[g] = S_{\rm g}^{\rm ren}[g] - S_{\rm div}[g] ,
\label{ga1}
\end{equation}
where the finite renormalized gravitational action is given by $S_{\rm g}^{\rm ren}[g] = S_{\rm EH}[g] + S_{\rm c}[g]$. The first term is the Einstein-Hilbert action
\begin{equation}
S_{\rm EH}[g]=\frac{1}{16\pi l_{\rm P}^2}\int d^4 x\sqrt{-g}\left(R-2\Lambda\right),
\label{r2}
\end{equation}
where $\Lambda$ is the cosmological constant and $l_{\rm P}$ is the Planck length, $l_{\rm P}=\sqrt {G}$, with $G$ being the gravitational constant. On the other hand, $S_{\rm c}[g]$ corresponds to the additional counterterms, which are quadratic in the curvature, and is given by
\begin{equation}
S_{\rm c}[g]=\int d^4 x\sqrt{-g}\left[\alpha\left(R_{\mu\nu\alpha\beta} R^{\mu\nu\alpha\beta}- R_{\mu\nu} R^{\mu\nu}\right)+\beta R^2\right],
\label{r3}
\end{equation}
where $\alpha$ and $\beta$ are dimensionless parameters. Note that since the bare parameters should be independent of the renormalization scale $\mu$, the renormalized parameters $1/G$, $\Lambda/G$, $\alpha$ and $\beta$ depend on $\mu$ in such a way that they cancel the $\mu$ dependence of the finite terms that arise when taking the $n \to 4$ limit in (\ref{r1}). More specifically, under a change the renormalization scale $\mu \to \mu'$ each one of the renormalized parameters changes by a term proportional to $\ln (\mu'/\mu)$. This also implies that the total effective action $\Gamma_\mathrm{CTP} [g^+,g^-]$ is invariant under the RG (i.e. independent of $\mu$).
Taking (\ref{ga1}) into account, the effective action in (\ref{ea1}) can be written as $\Gamma_\mathrm{CTP} [g^+,g^-] = S_{\rm g}^{\rm ren}[g^+] - S_{\rm g}^{\rm ren}[g^-]
+ \bar{S}_{\rm IF}^{\rm ren}[g^+,g^-]$, which is manifestly finite and where $\bar{S}_{\rm IF}^{\rm ren}[g^+,g^-] = S_{\rm IF}[g^+,g^-] - S_{\rm div}[g^+] + S_{\rm div}[g^-]$. Moreover, to make some of our expressions below more compact, we will reabsorb the part of the renormalized gravitational action which is quadratic in the curvature into the renormalized influence action so that we have
\begin{equation}
S_{\rm IF}^{\rm ren}[g^+,g^-]=S_{\rm IF}[g^+,g^-]-S_{\rm div}[g^+]+S_{\rm div}[g^-]+S_{\rm c}[g^+]-S_{\rm c}[g^-] ,
\label{r4}
\end{equation}
and the total effective action becomes
\begin{equation}
\Gamma_\mathrm{CTP} [g^+,g^-] = S_{\rm EH}[g^+] - S_{\rm EH}[g^-]
+ S_{\rm IF}^{\rm ren}[g^+,g^-] .
\label{ea2}
\end{equation}

Let us specialize these general results to the particular case that we are considering. For a metric of the form (\ref{2.1}), one has
\begin{eqnarray}
\fl
\int d^nx\sqrt{-g}\left(R_{\mu\nu\rho\sigma}R^{\mu\nu\rho\sigma}-R_{\mu\nu}R^{\mu\nu}\right)&=&
-{\cal V}(n-4)\int d\eta \left[ 3\left(\frac{\ddot a}{a}\right)^2-
\left(\frac{\dot a}{a}\right)^4\right]\nonumber\\
&&+ O\left((n-4)^2\right).
\label{r5}
\end{eqnarray}
where we neglected a total divergence that does not contribute to the equation of motion.
Using (\ref{2.2b}) and the fact that $\sqrt{-g} = a^n = a^4 \left(1 + (n-4) \ln a \right) + O\left( (n-4)^2 \right)$ one can proceed analogously for the terms in the integrand of (23) which are proportional to $R^2$, $R$ and a constant.
Substituting into (\ref{r4}) and taking into account the definition of the time-dependent mass below equation (\ref{2.3}), we obtain the following result for the renormalized influence action of our scalar field model:
\begin{eqnarray}
S_{\rm IF}^{\rm ren}[\bar a^+,\bar a^-] = S_{\rm IF}[\bar a^+,\bar a^-]
+\frac{\cal V}{32\pi^2}\,\Delta\!\int d\eta
&& \Bigg\{\beta\left(\frac{\ddot{\bar a}}{\bar a}\right)^2
+\frac{1}{90}\left(\frac{\dot{\bar a}}{\bar a}\right)^4 \nonumber \\
&& \ +\bar M^4\left[\frac{1}{n-4}+\ln (\bar a\mu)\right]\Bigg\},
\label{r6}
\end{eqnarray}
where we have introduced the difference notation, $\Delta f\equiv f^+-f^-$, and some numerical factors have been absorbed into the free parameter $\beta$. The $O(n-4)$ terms have been neglected, as they vanish in the limit $n\to 4$. Note that since the right-hand side of (\ref{r5}) vanishes when $n \to 4$, our renormalized influence action is independent of the renormalized parameter $\alpha$. On the other hand, the logarithmic dependence on $\mu$ is compensated by the $\mu$ dependence of $\beta$ as well as $1/G$ and $\Lambda/G$ in the Einstein-Hilbert action so that the total effective action is independent of $\mu$, as explained earlier. Moreover, although the argument of the logarithm in (\ref{r6}) has dimensions of mass, everything becomes well-defined when this term is combined with the bare influence action, as we will see below. From this renormalized influence action, one can obtain a finite expectation value of the stress-energy tensor by following the steps described in the previous subsection.

We close this subsection with a few remarks about the renormalized gravitational coupling constant $G(\mu)$, the renormalization scale $\mu$ and the experimental value of $G$ measured with a Cavendish-type torsion balance. In this sense, an important question that needs to be addressed is how the result of this kind of experiment can be used to fix the value of $G(\mu)$ at some scale $\mu$ and what is the most natural scale to consider. The Cavendish experiment can be essentially understood as a two-body scattering problem in the infrared limit. This is possible provided that there is a large separation of scales between the typical radius of curvature of the cosmological background (and in practice other local gravitational fields) and the Compton wavelength of the lightest massive particle: we need the typical scale of the experiment to be much larger than that Compton wavelength while being able to subtract or neglect other contributions to the local gravitational field. The condition on the Compton wavelength of the lightest massive particle guarantees that we are in the decoupling limit for all the massive fields, whereas massless fields do not contribute to the renormalization of $1/G$ and their correction to the Newtonian gravitational potential is suppressed by a factor $(l_\mathrm{P}/r)^2$, with $r$ being the distance between the two masses (the typical scale of the experiment) \cite{donoghue97,burgess04}. In that case, one can see that for a renormalization scale of the order of the mass of the lightest massive particle, $\mu \sim m$, $G(\mu)$ corresponds to the measured value if there are no additional massive fields. Otherwise, the measured value of $1/G$ would correspond to $1/G(m)$ plus a contribution of order $m_\mathrm{h}^2 \ln (m/m_\mathrm{h})$ for each heavier field with mass $m_\mathrm{h}$. Note that unless the mass $m$ of a field is not too far from the Planck mass, its contribution to the running of $1/G(\mu)$, which is of order $m^2 \ln(\mu/\mu')$ is extremely small compared to the value of $1/G(\mu)$ itself.
On the other hand, the value of the cosmological constant, which is much more sensitive to the running due to massive fields (even light ones), should also be determined in a similar way from cosmological observations. Finally, it is much harder to place upper bounds on the values of $\alpha$ and $\beta$ because the corrections from the terms quadratic in the curvature are very suppressed for phenomena with typical scales much larger than the Planck length.

\subsection{The semiclassical Einstein equations}
\label{sec2.4}

The semiclassical Einstein equation can be derived from the CTP effective action, as given by (\ref{ea2}), by functionally differentiating with respect to $g^+$ and then taking $g^+=g^-=g$. The Einstein-Hilbert action gives the Einstein tensor plus a cosmological constant term, and the renormalized influence action gives the renormalized expectation value of the stress tensor operator \cite{martin99b}:
\begin{equation}
G_{\mu\nu}+\Lambda g_{\mu\nu}=8\pi l_{\rm P}^2\langle T_{\mu\nu}\rangle_\Psi,
\label{2.23}
\end{equation}
where $G_{\mu\nu}$ is the Einstein tensor. Once the expectation value of the stress tensor is known, the backreaction of the quantum field on the mean geometry can be computed by substituting it into the semiclassical Einstein equation.
Because of spatial homogeneity and isotropy, only the equations for two components are independent. They can be chosen to be, for instance, the $00$ component and the stress tensor conservation law. Since the latter has already been taken into account when deriving the energy density from the trace, we only need the $00$ component, which is the semiclassical analog of the Friedmann equation, and in four dimensions has the form
\begin{equation}
 \dot a^2=H^2a^4+\frac{8\pi l_p^2}{3}\,a^2\langle T_{00}\rangle_\Psi,
\label{2.24}
\end{equation}
where we have introduced the Hubble constant $H=\sqrt{\Lambda/3}$. Note that, since we only know $\langle T_{00}\rangle_\Psi$ for $\eta>\eta_{\rm i}$, this equation can only be solved for $\eta>\eta_{\rm i}$. Two time-scales appear in the equation: the Hubble time $H^{-1}$ and the Planck time $t_{\rm P}=l_{\rm P}$. In order for the semiclassical approximation to be valid, these two time-scales have to be well separated, i.e. we need
$l_{\rm P} H \ll 1$.

As we will see below, the expectation value of the energy density contains up to third order time derivatives of the scale factor. Therefore, equation (\ref{2.24}) is a third order (integro-)differential equation, and its space of solutions is much larger than the one corresponding to the classical Friedmann equation. However, many of these solutions
have characteristic time-scales of the order of the Planck length and lie beyond the domain of validity of our low-energy EFT approach. In the spirit of this EFT approach, we  will look for perturbative solutions in powers of $(Hl_{\rm P})^2$:
\begin{equation}
a(\eta) = a_0(\eta) + (l_\mathrm{p} H)^2 a_1(\eta)+O\left( (l_{\rm P} H)^4 \right).
\label{2.25}
\end{equation}
A perturbative expansion may sometimes miss the right
  long-time behavior of the semiclassical solution. This can happen
  when the effect of the quantum corrections, although locally small,
  builds up over long times giving rise to substantial deviations from
  the classical solution. An example of such a situation is the
  evolution of a black hole spacetime when the back reaction of the
  emitted Hawking radiation is taken into account. One possibility in
  those cases is to modify the backreaction equation using an
  order-reduction procedure and then solve the resulting equation
  non-perturbatively~\cite{flanagan96,hu07b}.
If these cumulative effects were present in our case, the correction $a_1$ would get large and the perturbative approximation (\ref{2.25}) would no longer be valid. However, our perturbative treatment would still be useful to signal that these cumulative effects are taking place. In fact, one can check the validity of the perturbative approximation a posteriori by making sure that the perturbation remains small at all times.

Substituting (\ref{2.25}) into the Friedmann equation (\ref{2.24}) and solving order by order in perturbation theory, we get to lowest order that $a_0$ is just the scale factor of de Sitter spacetime,
\begin{equation}
a_0(\eta)=a_{\rm dS}(\eta)=-\frac{1}{H\eta},
\label{2.26}
\end{equation}
with $\eta<0$. The first order equation is
\begin{equation}
\dot a_1=-\frac{2}{\eta}a_1+\frac{4\pi}{3H^3}\langle T_{00}\rangle_{\Psi}^{\rm dS},
\label{2.27}
\end{equation}
where $\langle T_{00}\rangle_{\Psi}^{\rm dS}$ is the expectation value of the energy density evaluated on the de Sitter background. This is a first-order linear differential equation whose solution is readily obtained:
\begin{equation}
a_1(\eta)=\frac{4\pi}{3H^3\eta^2}\int^\eta d\eta'\eta'^2\langle T_{00}(\eta')\rangle_{\Psi}^{\rm dS}.
\label{2.28}
\end{equation}
In equations (\ref{2.26}) and (\ref{2.28}) we have dropped an arbitrary integration constant, because it can be set to zero by an appropriate choice of the origin of conformal time (see \cite{perez-nadal07} for details).

In the next section we will compute the expectation value of the energy density for non-conformal fields in a general RW background according to the steps described in this section, and then we will specialize it to a de Sitter background to obtain the first-order semiclassical correction to the de Sitter spacetime from equation (\ref{2.28}).

\section{Weakly non-conformal fields}
\label{sec3}

In this section we will obtain an explicit result for the influence action $S_{\rm{IF}}[\bar a^+,\bar a^-]$, making use of the formal expression (\ref{2.18b}) and the assumption that the field is weakly non-conformal, that is, $\bar M^2/\bar a^2\ll H^2$. The expectation value of the stress tensor in the state $|\Psi\rangle$ will then be derived and the semiclassical equation solved for the particular cases $(\nu=0,m\neq0)$ and $(\nu\neq0,m=0)$. 

For a weakly non-conformal field, the matrix $G$ in equation (\ref{2.18b}) can be computed perturbatively in powers of $\bar M^2$. Following \cite{campos94,campos97,calzetta97c} we define $A=A^0+V$, where
the matrix $V$ includes the time-dependent interaction with
$V_{+}=-\bar M^{2}_+$, and $V_{-}=\bar M^2_-$. Then, up to second order in $\bar M^2$, 
\begin{equation}
G=G^0(1-VG^0+VG^0VG^0+\dots\,),
\label{3.1}
\end{equation}
where $G^0=(A^{0})^{-1}$ is the Minkowski, massless CTP propagator.
Its components are $G^0_{++}=\Delta_F$,
$G^0_{--}=-\Delta_D$, $G^0_{+-}=-\Delta^+$ and $G^0_{-+}=\Delta^-$,
where $\Delta_F$ and $\Delta_D$ are respectively the Feynman and
Dyson propagators and $\Delta^\pm$ are the Wightman functions:
\begin{eqnarray}
&&\Delta_{F/D}(x)=-\int\frac{d^n p}{(2\pi)^n}\frac{e^{ip\cdot x}}{p^2
\mp i\epsilon},\nonumber\\
&&\Delta^{\pm}(x)=\pm 2\pi i \int\frac{d^n p}{(2\pi)^n}e^{ip\cdot x}\delta(p^2)\theta(\mp p^0).
\label{3.2}
\end{eqnarray}
Substituting into equation (\ref{2.18b}) we have (up to second order in
$\bar M^2$)
\begin{eqnarray}
S_{\rm{IF}}[\bar a^+,\bar a^-]&=&-\frac{i}{2}\textrm{tr}\ln G^0- \frac{i}{2}\textrm{tr} (\bar M_{+}^2\Delta_F)- \frac{i}{2}\textrm{tr} (\bar M_{-}^2\Delta_D)
\nonumber\\&&-\frac{i}{4}\textrm{tr} (\bar M_{+}^2\Delta_F\bar M_{+}^2 \Delta_F)-\frac{i}{4}\textrm{tr} (\bar M_{-}^2\Delta_D\bar M_{-}^2\Delta_D)\nonumber\\
&&- \frac{i}{2}\textrm{tr} (\bar M_{+}^2\Delta^+ \bar M_{-}^2\Delta^-).
\label{3.3}
\end{eqnarray}
The first three terms in this equation do not contribute to the expectation value of the stress tensor. The first one is independent of $\bar a$, whereas the second and third terms are tadpoles which are identically zero in dimensional regularization \cite{leibbrandt75}. Therefore, the influence action is quadratic in $\bar M^2$. Computing the last three terms and expanding in powers of $(n-4)$, we get
\begin{eqnarray}
\fl
\mathrm{Re}\, S_{\rm IF}
= -\frac{{\cal V}}{32\pi^2}\int d\eta\left\{\frac{1}{n-4}\Delta \bar M^4 (\eta)
+2\Delta \bar M^2 (\eta) \int d\eta' H(\eta-\eta';\theta)\Sigma \bar M^2(\eta')\right\},
\label{3.4}
\end{eqnarray}
where the terms of order $O(n-4)$ have been neglected. The range of both $\eta$ and $\eta'$ integrations is $(-\infty,\eta_f]$, where $\eta_f$ is an arbitrary final time (larger than any other time that we may be interested in). We have used the difference and semisum notations, $\Delta
f\equiv f^+-f^-$ and $\Sigma f \equiv(f^+ +f^-)/2$ respectively.
As shown in \cite{campos97}, the kernel $H(\eta-\eta^\prime;\theta)$ is given by
\begin{equation}
H(\eta-\eta^\prime;\theta)= \int \frac{d\omega}{2\pi}e^{-i\omega (\eta-\eta^\prime)}\left(\ln \frac{|\omega|}{\theta} +\frac{i\pi}{2}\textrm{sign}(-\omega)  \right),
\label{3.5}
\end{equation}
where $\theta^2=\exp(2+\ln 4\pi -\gamma)$, and $\gamma$ is the Euler-Mascheroni constant. In equation (\ref{3.4}) we have dropped an imaginary term, that does not contribute to the expectation value of the stress tensor because it is proportional to $(\Delta\bar M^2)^2$. This is consistent with the fact that the expectation value of an observable must be a real number. Using equation (\ref{r6}) and taking the limit $n\to 4$, we obtain the renormalized influence action:
\begin{eqnarray}
\fl
\mathrm{Re}\, S_{\rm IF}^{\rm ren}
= \frac{\cal V}{32\pi^2} \! \int \!\! d\eta
&& \Bigg\{\beta\Delta\left(\frac{\ddot{\bar a}(\eta)}{\bar a(\eta)}\right)^2
+ \frac{1}{90}\Delta\left(\frac{\dot{\bar a}(\eta)}{\bar a (\eta)}\right)^4\nonumber\\
&& \ +\Delta\left(\bar M^4(\eta) \ln\bar a(\eta)\right)
- 2\Delta\bar M^2(\eta) \int d\eta' H(\eta-\eta';\mu)\Sigma\bar M^2(\eta')\Bigg\},
\label{3.6}
\end{eqnarray}
where the numerical factor $\theta$ has been absorbed into $\mu$. Notice that as has been anticipated in subsection \ref{sec2.3}, the term involving $\ln\mu$ in (\ref{r6}) combines with the term with $\ln |\omega|$ in (\ref{3.5}), rendering the argument of the logarithm dimensionless.

Let us finally turn our attention to the non-local part of the influence action. The Fourier transform (\ref{3.5}) can be explicitly computed. If we define the functional
\begin{equation}
\kappa[f;\eta)=\int_{-\infty}^{\eta_f}d\eta'H(\eta-\eta';\mu)f(\eta'),
\label{3.7}
\end{equation}
using equation~(VII.7.18) in \cite{schwartz66}, it can be shown to be
\begin{equation}
\kappa[f;\eta)=-\lim_{\epsilon\to 0^+}\left\{\int^{\eta-\epsilon}_{-\infty} \frac{d\eta^\prime}{\eta-\eta^\prime} f(\eta^\prime)+(\ln\epsilon+\ln{\mu}+\gamma)f(\eta)\right\}.
\label{3.8}
\end{equation}
In order for the integral in this equation to be finite in its lower limit, $f$ must vanish at past infinity. Assuming that this condition is satisfied, the functional $\kappa[f;\eta)$ is finite at $\eta$ provided that $f$ is continuous at $\eta$. Indeed, in this case near $\eta'=\eta$ we can write $f(\eta')=f(\eta)+O(\eta-\eta')$, so that the integral in (\ref{3.8}) has the form
\begin{equation}
\int_{-\infty}^{\eta-\epsilon}\frac{d\eta'}{\eta-\eta'}f(\eta')=-f(\eta)\ln\epsilon+{\textrm{finite terms}},
\label{3.9}
\end{equation}
and the contributions that diverge in the limit $\epsilon\to 0^+$ cancel. Therefore, in order for the functional $\kappa[f;\eta)$ to be well-defined, the function $f$ must be continuous and vanish when $\eta \to -\infty$.  On the other hand, for a function satisfying these conditions it is easy to see that
\begin{equation}
\frac {d}{d\eta}\kappa[f;\eta)=\kappa[\dot f;\eta).
\label{3.10}
\end{equation}
The expectation value of the trace of the stress tensor, which will be obtained by functional differentiation of the influence action (\ref{3.6}), will include a term proportional to $\kappa[\bar M^2;\eta)$, and in some cases there will also be terms proportional to its first and second derivatives, as will be shown below. The condition (\ref{2.7}) already ensures that $\bar M^2$ vanishes at past infinity. Now we see that $\bar M^2$ must also satisfy some ``smoothness'' conditions. In particular, in order for the trace not to diverge at the initial time $\eta_{\rm i}$, the scale factors $a_\Psi$ and $a$ in equation (\ref{2.11}) must be matched smoothly enough. This corresponds to choosing regular initial states, namely, adiabatic vacua of the appropriate order, as is explained in the appendix.

\subsection{Massive conformally coupled fields}
\label{sec3.1}

Here we restrict our attention to the particular case of conformally coupled fields (with $\nu=0$), but keeping $m\ne 0$. In this case, the condition (\ref{2.7}) implies that $a_\Psi$ vanishes at past infinity. The renormalized influence action (\ref{3.6}) takes the form
\begin{eqnarray}
\fl
\mathrm{Re}\, S_{\rm IF} = \frac{\cal V}{32\pi^2}\int d\eta\Bigg\{\beta\Delta\left(\frac{\ddot{\bar a}}{\bar a}\right)^2+\frac{1}{90}\Delta\left(\frac{\dot{\bar a}}{\bar a}\right)^4
+m^4\left[\Delta\left(\bar a^4\ln\bar a\right)-2\Delta\bar a^2\kappa[\Sigma\bar a^2]\right]\Bigg\},
\label{3.11}
\end{eqnarray}
where we have dropped the superscript ``ren'' for simplicity, and $\kappa$ is the functional defined in equations (\ref{3.7}) and (\ref{3.8}). Functionally differentiating with respect to $\bar a^+$ and making use of equation (\ref{2.19}), we obtain the expectation value of the trace of the stress tensor, which after some manipulations reads
\begin{eqnarray}
\bar a^3\dot{\bar a}\langle \bar T^{\mu}_\mu\rangle = -\frac{d}{d\eta} \, && \Bigg\{\beta\Bigg[\dot{\bar a}\frac{d}{d\eta}\Bigg(\frac{\ddot{\bar a}}{\bar a^2}\Bigg)-\frac{1}{2}\Bigg(\frac{\ddot{\bar a}}{\bar a}\Bigg)^2\Bigg]+\frac{1}{960\pi^2}\Bigg(\frac{\dot{\bar a}}{\bar a}\Bigg)^4\nonumber\\
&& \ -\frac{m^4}{32\pi^2}\Big[\bar a^4\ln\bar a+F[\bar a^2]\Big]\Bigg\},
\label{3.12}
\end{eqnarray}
where we again absorbed some numerical factors into the free parameter $\beta$, and the functional $F$ is defined as
\begin{equation}
F[\bar a^2;\eta)=-2\int_{-\infty}^\eta d\eta'\frac{d\bar a^2}{d\eta'}\kappa[\bar a^2;\eta').
\label{3.13}
\end{equation}
We see that the expectation value of the trace of the stress tensor involves a term proportional to $\kappa[\bar a^2]$. According to the discussion below (\ref{3.8}), in order for the trace to be finite at the initial time $\eta_{\rm i}$, the auxiliary scale factor $a_\Psi$ employed in the construction of the vacuum $|\Psi\rangle$ must verify
\begin{equation}
a_\Psi(\eta_{\rm i})=a(\eta_{\rm i}).
\label{3.14}
\end{equation}
Using equations (\ref{2.21}) and (\ref{2.22}) it is straightforward to obtain the expectation value of the energy density for $\eta>\eta_{\rm i}$ given a scale factor $a(\eta)$ and an initial state $|\Psi\rangle$:
\begin{eqnarray}
a^2\langle T_{00}\rangle_\Psi &=& \beta\Bigg[\dot{a}\frac{d}{d\eta}\Bigg(\frac{\ddot{a}}{a^2}\Bigg)-\frac{1}{2}\Bigg(\frac{\ddot{a}}{ a}\Bigg)^2\Bigg]+\frac{1}{960\pi^2}\Bigg(\frac{\dot{a}}{a}\Bigg)^4 \nonumber \\
&&-\frac{m^4}{32\pi^2}\Big[a^4\ln a+F[\bar a^2]\Big].
\label{3.15}
\end{eqnarray}
This result is valid for any scale factor $a(\eta)$ and for any vacuum $|\Psi\rangle$ (as defined in subsection~\ref{sec2.1}) satisfying the condition (\ref{3.14}). Let us now particularize this to the case of a de Sitter background. Since de Sitter spacetime has constant curvature, the first term in (\ref{3.15}) vanishes: its origin is the term proportional to $R^2$ in the influence action, which yields a term proportional to $\Box R$ after functional differentiation. Therefore, there is no dependence on the free parameter $\beta$, and one is left with
\begin{equation}
a^2_{\rm dS}\langle T_{00}\rangle_{\Psi}^{\rm dS}=\frac{1}{960\pi^2}\frac{1}{(-\eta)^4}-\frac{m^4}{32\pi^2}\left[\frac{-1}{(-H\eta)^4}\ln (-H\eta)+F[\bar a^{2}_0;\eta)\right],
\label{3.16}
\end{equation}
where $\bar a^{2}_0(\eta)$ is defined as
\begin{equation}
\bar a_{0}^2(\eta)=\left\{\begin{array}{ll}
a_{\rm dS}^2 & \textrm{for $\eta>\eta_{\rm i}$}\\
a_{\Psi}^2=a_{\rm dS}^2+\delta a_{\Psi}^2 & \textrm{for $\eta\le\eta_{\rm i}$}.
\end{array}\right.
\label{3.17}
\end{equation}
Note that, as a consequence of the conditions (\ref{2.7}) and (\ref{3.14}), we must have
\begin{equation}
\lim_{\eta\to -\infty}\delta a_{\Psi}^2=0,\qquad \delta a_{\Psi}^2 (\eta_{\rm i})=0.
\label{3.18}
\end{equation}
One possibility is, of course, $\delta a_{\Psi}^2 (\eta)=0 \;\, \forall \, \eta \leq \eta_{\rm i}$; in this case, the state is the \emph{in}-vacuum of de Sitter spacetime, which is the so-called Bunch-Davies vacuum. Substituting the scale factor (\ref{3.17}) into equation (\ref{3.13}), we get after some manipulations that the state-dependent part of the expectation value of the energy density can be written as
\begin{eqnarray}
F[\bar a^{2}_0;\eta) &=& F[a^{2}_{\rm dS};\eta)+F[\delta a^{2}_\Psi;\eta_{\rm i})
\nonumber \\
&&-2\int_{-\infty}^{\eta_{\rm i}}d\eta'\delta a_{\Psi}^2(\eta')\,{\cal P}\!\int_{-\infty}^{\eta}\frac{d\eta''}{\eta'-\eta''}\frac{da_{\rm dS}^2}{d\eta''},
\label{3.19}
\end{eqnarray}
where ${\cal P}$ denotes Cauchy's principal value. The first term in this equation, as well as the second integral in the third term, can be explicitly computed. The latter involves a logarithm that can be expanded in a Taylor series. The result is
\begin{eqnarray}
F[\bar a^{2}_0;\eta) &=& \frac{1}{(-H\eta)^4}\left[\ln(-\mu\eta)+\gamma-\frac{3}{4}\right]
\nonumber \\
&&-\sum_{n=0}^{\infty}A_{n}^\Psi(-H\eta)^{n-2}-B_\Psi\ln(-H\eta),
\label{3.20}
\end{eqnarray}
where the state-dependent, dimensionless constants $A_{n}^\Psi$ and $B_\Psi$ are given by
\begin{eqnarray}
A_{n}^\Psi&=&\frac{4}{H^n(n-2)}\int_{-\infty}^{\eta_{\rm i}}\frac{d\eta'}{(-\eta')^{n+1}}\delta a_{\Psi}^2(\eta')\qquad {\textrm{for }}n\ne 2\nonumber\\
A_{2}^\Psi&=&-\frac{4}{H^2}\int_{-\infty}^{\eta_{\rm i}}\frac{d\eta'}{(-\eta')^3}\ln(-H\eta')\delta a_{\Psi}^2(\eta')-F[\delta a^{2}_\Psi;\eta_{\rm i})\nonumber\\
B_\Psi&=&\frac{4}{H^2}\int_{-\infty}^{\eta_{\rm i}}\frac{d\eta'}{(-\eta')^3}\delta a_{\Psi}^2(\eta').
\label{3.21}
\end{eqnarray}
All these integrals are finite due to the first of the conditions (\ref{3.18}). Note also that they all vanish when the state is the Bunch-Davies vacuum. Substituting the result (\ref{3.20}) into equation (\ref{3.16}) we obtain an explicit result for the expectation value of the energy density of the scalar field in de Sitter spacetime for the state $|\Psi\rangle$: 
\begin{equation}
\langle T_{00}\rangle_{\Psi}^{\rm dS}=\langle T_{00}\rangle_{\rm BD}^{\rm dS}+\frac{m^4}{32\pi^2}\left[\sum_{n=0}^{\infty}A_{n}^\Psi(-H\eta)^{n}+B_\Psi(-H\eta)^2\ln(-H\eta)\right],
\label{3.22}
\end{equation}
where $\langle T_{00}\rangle_{\rm BD}^{\rm dS}$ is the same expectation value in the Bunch-Davies vacuum,
\begin{equation}
\langle T_{00}\rangle_{\rm BD}^{\rm dS}=\frac{1}{(-H\eta)^2}\left[\frac{H^4}{960\pi^2}-\frac{m^4}{32\pi^2}\left(\ln\frac{\mu}{H}+\gamma-\frac{3}{4}\right)\right].
\label{3.23}
\end{equation}
This result agrees through order $m^4$ with the exact result in equation (6.183) of \cite{birrell94}. To show the equivalence one needs to take into account that their renormalization prescription corresponds to taking $\mu$ equal to $m$ times a dimensionless number involving the Euler-Mascheroni constant, and the fact that their counterterms are given by (\ref{r1}) but with $(\xi - 1/6)$ instead of $\nu = (\xi - \xi_\mathrm{c}) = (\xi - 1/6) - (n-4)/36 + O((n-4)^2)$.
Note also that the $\mu$-dependent term in (\ref{3.23}) is exactly compensated by the $\mu$ dependence of the cosmological constant in the renormalized Einstein-Hilbert action.%
\footnote{There is no $\mu$ dependence associated with the term proportional to the Ricci scalar because there is no divergence associated with that term for $\nu=0$, as can be seen from (\ref{r1}). However, for the massive and non-conformally coupled case there would be such a dependence.}
Moreover, there is no $\mu$ dependence in the state-dependent part of (\ref{3.22}), as can be seen from examination of the coefficients (\ref{3.21}). The only one that could potentially depend on $\mu$ is $A_{2}^\Psi$, but it actually does not: the $\mu$-dependent part of $\kappa[f;\eta)$ is local, as can be seen from its definition (\ref{3.8}), so that the $\mu$-dependent part of $F[\delta a^{2}_\Psi;\eta_{\rm i})$ is proportional to
\begin{equation*}
2\int_{-\infty}^{\eta_{\rm i}}d\eta'\delta a^{2}_\Psi \frac{d(\delta a^{2}_\Psi)}{d \eta'}=(\delta a^{2}_\Psi)^2(\eta_{\rm i})=0,
\end{equation*}
as follows from conditions (\ref{3.18}).
Therefore, the semiclassical Einstein equation is $\mu$-independent, as emphasized in subsection \ref{sec2.3}.

Another point that one must check is that the infinite series in (\ref{3.22}) converges. To see that this is indeed the case, note that the absolute value of the coefficients $A_{n}^\Psi$ in (\ref{3.21}) (for $n>2$) is bounded from above as follows:
\begin{equation}
|A_{n}^\Psi|\le 4\,{\rm{max}}(|\delta a^{2}_\Psi|)\frac{1}{n(n-2)}\frac{1}{(-H\eta_{\rm i})^n}.
\label{3.24}
\end{equation}
Therefore, the absolute value of the $n$-th term of the series is
\begin{equation}
|A_{n}^\Psi|(-H\eta)^n\le 4\,{\rm{max}}(|\delta a^{2}_\Psi|)\frac{1}{n(n-2)}\left(\frac{\eta}{\eta_{\rm i}}\right)^n\le4\,{\rm{max}}(|\delta a^{2}_\Psi|)\frac{1}{n(n-2)},
\label{3.25}
\end{equation}
where the last inequality holds for $\eta\ge\eta_{\rm i}$ (recall that both $\eta$ and $\eta_{\rm i}$ are negative). Now, the infinite series in (\ref{3.22}) converges because the series $\sum 1/n^2$, which is Riemann's zeta function $\zeta(2)$, is finite:
\begin{equation}
\sum_{n=1}^\infty\frac{1}{n^2}= \zeta(2)=\frac{\pi^2}{6}.
\label{3.26}
\end{equation}
As a final remark, notice that the expectation value (\ref{3.22}) tends to the Bunch-Davies one (\ref{3.23}) at future infinity, $\eta\to 0$, regardless of the state $|\Psi\rangle$. 

The final step is to use the result (\ref{3.22}) to compute the semiclassical correction to the scale factor of de Sitter spacetime using equation (\ref{2.28}). The result is
\begin{equation}
a_1(\eta)=a_{1}^{\rm BD}(\eta)-\frac{m^4}{24\pi H^4}\left[\sum_{n=1}^\infty C_{n}^\Psi(-H\eta)^n+D_\Psi(-H\eta)^3\ln(-H\eta)\right],
\label{3.27}
\end{equation}
where $a_{1}^{\rm BD}(\eta)$ is the correction associated with the Bunch-Davies vacuum,
\begin{equation}
a_{1}^{\rm BD}(\eta)=\frac{1}{H\eta}\left[\frac{1}{720\pi}-\frac{m^4}{24\pi H^4}\left(\ln\frac{\mu}{H}+\gamma-\frac{3}{4}\right)\right],
\label{3.28}
\end{equation}
and the state-dependent coefficients $C_{n}^\Psi$ and $D_\Psi$ are
\begin{eqnarray}
C_{n}^\Psi&=&\frac{A_{n-1}^\Psi}{n+2}\qquad{\textrm{for }}n\ne 3\nonumber\\
C_{3}^\Psi&=&\frac{A_{2}^\Psi}{5}-\frac{B_\Psi}{25}\nonumber\\
D_\Psi&=&\frac{B_\Psi}{5}.
\label{3.29}
\end{eqnarray}
Substituting this result into the expansion (\ref{2.25}), we finally conclude that the solution of the semiclassical Friedmann equation (\ref{2.24}) is
\begin{eqnarray}
a(\eta)&=&-\frac{1}{\tilde H\eta}-(l_{\rm P}H)^2\frac{m^4}{24\pi H^4}\left[\sum_{n=1}^\infty C_{n}^\Psi(-H\eta)^n+D_\Psi(-H\eta)^3\ln(-H\eta)\right]\nonumber\\
&&+O\left((l_{\rm P}H)^4\right),
\label{3.30}
\end{eqnarray}
where the corrected Hubble constant $\tilde H$ is given by
\begin{equation}
\tilde H=H\left\{1+(l_{\rm P}H)^2\left[\frac{1}{720\pi}-\frac{m^4}{24\pi H^4}\left(\ln\frac{\mu}{H}+\gamma-\frac{3}{4}\right)\right]\right\}.
\label{3.31}
\end{equation}
Note that the explicitly $\mu$-dependent term in (\ref{3.31}) is compensated by the $\mu$ dependence of $H$, which it inherits form the $\mu$ dependence of $\Lambda$, so that $\tilde H$ is $\mu$-independent. [The $\mu$ dependence of $H$ in the terms of order $(l_{\rm P}H)^2$ is irrelevant since it would give rise to contributions of higher order in $(l_{\rm P}H)^2$.]
If the state chosen is the Bunch-Davies vacuum, then the coefficients $C_{n}^\Psi$ and $D_\Psi$ vanish, and the solution (\ref{3.30}) of the semiclassical Friedmann equation is de Sitter spacetime with the corrected Hubble constant $\tilde H$. If we choose any other state, then the solution is no longer the scale factor of de Sitter spacetime, but tends to it at late times (as $\eta \to 0^-$). Finally, notice that since there is no dependence on the free renormalization parameters $\alpha$ and $\beta$, this result is fully predictive once the cosmological constant and the gravitational coupling constant (or, equivalently, $l_{\rm P}$) are measured.

It should be emphasized that the logarithmic term which appears explicitly in (\ref{3.16}) and comes from the local term proportional to $m^4$ and $\ln a$ in (\ref{3.11}) is cancelled out by the first term on the right-hand side of (\ref{3.20}), which contains the contributions from the non-local term in (\ref{3.11}). If it had not been cancelled out by the contribution from the non-local term, such a local term would have dominated the result for the stress tensor at late times, precluding the existence of a self-consistent de Sitter solution and altering our main conclusions about the semiclassical stability of de Sitter spacetime. This is an important point whose implications will be further discussed in section~\ref{sec5}.

\subsection{Massless non-conformally coupled fields}
\label{sec3.2}

Next we move to the opposite situation: we will consider massless fields but non-conformally coupled (i.e. $m=0$ but $\nu\ne 0$). We will proceed in strict analogy with the previous subsection, so some of the details will be skipped (this case has been considered in great detail in \cite{perez-nadal07}). In this case, $M(\eta)$ in four dimensions reads
\begin{equation}
M^2=6\nu\frac{\ddot a}{a}\equiv 6\nu u,
\label{3.32}
\end{equation}
as follows from equation (\ref{2.2b}). Condition (\ref{2.7}) translates here into $u_\Psi\to 0$ at past infinity. Specializing the influence action (\ref{3.6}) to this case, functionally differentiating it with respect to $\bar a^+$ and applying equation (\ref{2.19}), one finds that the expectation value of the trace of the stress tensor depends on the functional $\kappa[\bar u;\eta)$, as well as its first and second derivatives. Therefore, from the discussion around equations (\ref{3.9}) and (\ref{3.10}), we see that, in order to keep the trace finite at the initial time, the scale factor $a_\Psi$ employed to generate the state $|\Psi\rangle$ must satisfy the following conditions:
\begin{equation}
u_\Psi(\eta_{\rm i})=u(\eta_{\rm i})\qquad \dot u_\Psi(\eta_{\rm i})=\dot u(\eta_{\rm i})\qquad \ddot u_\Psi(\eta_{\rm i})=\ddot u(\eta_{\rm i}).
\label{3.33}
\end{equation}
In other words, the scale factors before and after the initial time must be matched with continuity up to the fourth derivative. Once the trace is known, the expectation value of the energy density of the field for $\eta>\eta_{\rm i}$, given an initial state $|\Psi\rangle$ and a scale factor $a(\eta)$, is obtained from equations (\ref{2.21}) and (\ref{2.22}). The result is
\begin{eqnarray}
a^2\langle T_{00}\rangle_\Psi&=&\beta\left[\dot a \frac{d}{d\eta}\left(\frac{\ddot a}{a^2}\right)-\frac{1}{2}\left(\frac{\ddot a}{a}\right)^2\right]+\frac{1}{960\pi^2}\left(\frac{\dot a}{a}\right)^4\nonumber\\&&-\frac{9\nu^2}{4\pi^2}\left\{\ln a\left[\dot a \frac{d}{d\eta}\left(\frac{\ddot a}{a^2}\right)-\frac{1}{2}\left(\frac{\ddot a}{a}\right)^2\right]+\frac{\dot{a}^2\ddot a}{a^3}+T[a,\bar u]\right\},
\label{3.34}
\end{eqnarray}
where the non-local, state-dependent part reads
\begin{equation}
T[a,\bar u;\eta)=\left(\frac{\dot a}{a}\right)^{\!2}\!\!\!(\eta)\kappa[\bar u;\eta)-\left(\frac{\dot a}{a}\right)\!\!(\eta)\kappa[\dot {\bar u};\eta)+\int^{\eta}_{-\infty}d\eta^\prime \bar u(\eta^\prime)\kappa[\dot{\bar u};\eta^\prime).
\label{3.35}
\end{equation}
Note that the first and second terms in equation (\ref{3.34}) are just the same as in (\ref{3.15}). They both come from the first two terms in the influence action (\ref{3.6}), which are independent of the particular form of the time-dependent mass $M(\eta)$. In order to specialize this result to a de Sitter background, but keeping the state $|\Psi\rangle$ general, one must compute $T[a_{\rm dS},\bar u_0;\eta)$ with
\begin{equation}
\bar u_{0}(\eta)=\left\{\begin{array}{ll}
u_{\rm dS} & \textrm{for $\eta>\eta_{\rm i}$}\\
u_{\Psi}=u_{\rm dS}+\delta u_{\Psi} & \textrm{for $\eta\le\eta_{\rm i}$},
\end{array}\right.
\label{3.36}
\end{equation}
where $u_{\rm dS}(\eta)=2/\eta^2$ [$u_{\Psi}(\eta)$ was denoted by $v(\eta)$ in \cite{perez-nadal07}]. Note that conditions (\ref{2.7}) and (\ref{3.33}) translate into the following conditions on $\delta u_\Psi$:
\begin{equation}
\lim_{\eta\to -\infty}\delta u_\Psi=0\qquad \delta u_\Psi(\eta_{\rm i})=\delta \dot u_\Psi(\eta_{\rm i})=\delta\ddot u_\Psi(\eta_{\rm i})=0.
\label{3.37}
\end{equation}
Once again, taking $\delta u_\Psi=0$ corresponds to the \emph{in}-vacuum for de Sitter spacetime, i.e. the Bunch-Davies vacuum. Substituting the scale factor $a_{\rm dS}$ and the function $\bar u_0$ into equation (\ref{3.35}), we get that the state-dependent part of the expectation value of the energy density in de Sitter spacetime is
\begin{equation}
T[a_{\rm dS},\bar u_0;\eta)=-H^4\left[\frac{3}{2}\frac{1}{(-H\eta)^4}+\sum_{n=0}^\infty P_{n}^\Psi(-H\eta)^{n-2}+Q_\Psi\ln(-H\eta)\right],
\label{3.38}
\end{equation}
where the dimensionless state-dependent constants $P_{n}^\Psi$ and $Q_\Psi$ are given by
\begin{eqnarray}
P_{n}^\Psi &=&\frac{n^2+n-2}{H^{n+2}(n-2)}\int_{-\infty}^{\eta_{\rm i}}\frac{d\eta'}{(-\eta')^{n+1}}\delta u_\Psi(\eta')\qquad{\textrm{for }}n\ne 2\nonumber\\
P_{2}^\Psi &=&-\frac{1}{H^4}\int_{-\infty}^{\eta_{\rm i}}d\eta'\delta u_\Psi(\eta')\left[\frac{4}{(-\eta')^3}\ln(-H\eta')-\frac{5}{(-\eta')^{3}}+\kappa[\delta \dot u_\Psi;\eta')\right]\nonumber\\
Q_\Psi &=&\frac{4}{H^4}\int_{-\infty}^{\eta_{\rm i}}\frac{d\eta'}{(-\eta')^{3}}\delta u_\Psi(\eta').
\label{3.39}
\end{eqnarray}
These state-dependent coefficients are finite, as a consequence of the first condition in (\ref{3.37}), and vanish in the particular case of the Bunch-Davies vacuum. Making use of this result, it is now straightforward to explicitly obtain the expectation value of the energy density for the state $|\Psi\rangle$ in de Sitter spacetime:
\begin{equation}
\langle T_{00}\rangle_{\Psi}^{\rm dS}=\langle T_{00}\rangle_{\rm BD}^{\rm dS}+\frac{9\nu^2H^4}{4\pi^2}\left[\sum_{n=0}^{\infty}P_{n}^\Psi(-H\eta)^{n}+Q_\Psi(-H\eta)^2\ln(-H\eta)\right],
\label{3.40}
\end{equation}
where the first term is the expectation value in the Bunch-Davies vacuum, and reads
\begin{equation}
\langle T_{00}\rangle_{\rm BD}^{\rm dS}=\frac{H^4}{(-H\eta)^2}\left(\frac{1}{960\pi^2}-\frac{9\nu^2}{8\pi^2}\right).
\label{3.41}
\end{equation}
In this case, there is no $\mu$ dependence at all. First, there is no dependence in the state-dependent part: the only coefficient which could depend on $\mu$ is $P_{2}^\Psi$ and one can show that it actually does not, exactly in the same way as done for $A_{2}^\Psi$ in the previous subsection. Second, there is no dependence on $\mu$ in the state-independent part either because the only term that depends on $\mu$ is proportional to $\Box R$, which vanishes when evaluated on a de Sitter background. This term is related to the $\beta R^2$ term of the renormalized gravitational action. On the other hand, there is no dependence on $\mu$ associated with the terms of the Einstein-Hilbert action because there are no divergences involving those terms in the massless case, as can be seen from (\ref{r1}).

The convergence of the infinite series in (\ref{3.40}) can be checked by following the same steps as in the previous subsection. One also needs to use the relation
\begin{equation*}
\int_{-\infty}^{\eta_{\rm i}}\frac{d\eta'}{(-\eta')^{n+1}}\delta u_\Psi(\eta')=\frac{1}{n(n-1)}\int_{-\infty}^{\eta_{\rm i}}\frac{d\eta'}{(-\eta')^{n-1}}\delta\ddot u_\Psi(\eta'),
\end{equation*}
which is a consequence of the conditions in (\ref{3.37}). Another interesting point is that, analogously to the massive conformally coupled case, the expectation value (\ref{3.40}) tends to the Bunch-Davies one in the asymptotic future (i.e. when $\eta\to 0^-$) regardless of the state $|\Psi\rangle$.

We can now compute the semiclassical correction to de Sitter spacetime, making use of the expectation value (\ref{3.40}) and equation (\ref{2.28}). Substituting the result into the expansion (\ref{2.25}), we obtain that the solution to the semiclassical Friedmann equation is 
\begin{eqnarray}
a(\eta) &=& -\frac{1}{\tilde H\eta}-(l_{\rm P}H)^2\frac{3\nu^2}{\pi}\left[\sum_{n=1}^\infty R_{n}^\Psi(-H\eta)^n+S_\Psi(-H\eta)^3\ln(-H\eta)\right] \nonumber \\
&&+O\left((l_{\rm P}H)^4\right),
\label{3.42}
\end{eqnarray}
where the relation between the coefficients in the stress tensor and the solution for the scale factor is the same as in the previous subsection:
\begin{eqnarray}
R_{n}^\Psi&=&\frac{P_{n-1}^\Psi}{n+2}\qquad{\textrm{for }}n\ne 3\nonumber\\
R_{3}^\Psi&=&\frac{P_{2}^\Psi}{5}-\frac{Q_\Psi}{25}\nonumber\\
S_\Psi&=&\frac{Q_\Psi}{5}.
\label{3.43}
\end{eqnarray}
In this case the corrected Hubble constant is
\begin{equation}
\tilde H=H\left[1+(l_\mathrm{p}H)^2 \left(\frac{1}{720\pi}
-\frac{3\nu^2}{2\pi}\right)\right].
\label{3.44}
\end{equation}
If the state chosen is the Bunch-Davies vacuum, the solution to the semiclassical Friedmann equation is just de Sitter spacetime. For any other state, the solution is the scale factor of de Sitter spacetime plus some terms, but these terms vanish at future infinity. The result (\ref{3.42}) is fully predictive, as there is no dependence on the free renormalization parameters $\alpha$ and $\beta$.

\section{Strongly non-conformal fields}
\label{sec4}

In this section we briefly describe how to obtain the influence action for strongly non-conformal fields corresponding to a large mass $m^2 \gg H^2$ (the curvature coupling parameter $\nu$ can take arbitrary values as long as $m^2 \gg \nu R$).

Whenever the mass of the fields is much larger than the inverse of the curvature radius, one can introduce a quasi-local expansion for the Feynman propagator and other related Green functions (see section 3.6 of \cite{birrell94}). This can then be used to calculate the influence action as a local expansion of terms involving positive powers of the curvature times the appropriate power of the mass squared which gives the right dimension, as explained in sections 6.1 and 6.2 of \cite{birrell94}. This kind of expansion is known as the Schwinger-DeWitt expansion and it is often employed within some covariant regularization scheme in order to identify the divergent contributions to the effective action, which can be dealt with by a suitable renormalization of local terms in the bare gravitational action. When identifying divergent terms in four-dimensional spacetimes, the expansion is truncated beyond order one and only includes terms of order $m^4$, $m^2$ and $1$. After subtraction of the divergences associated with those terms through the usual renormalization procedure, one is finally left with the same kind of terms multiplied by finite coupling constants. There are, however, additional finite contributions from terms in the expansion corresponding to negative powers of $m^2$. In particular, the terms of order $m^{-2}$
were explicitly considered in \cite{frolov84}, where the resulting effective action was finally used to calculate the expectation value of the stress-tensor operator for fields in a Kerr black hole spacetime.

In fact, using arguments based on the relevant symmetries (diffeomorphism invariance in this case) and power counting, one can easily infer the form of the influence action in this regime except for the precise values of dimensionless coefficients. This kind of approach is very common when dealing with effective theories and it is nicely illustrated by the example of the Euler-Heisenberg Lagrangian \cite{itzykson80}, which describes the effects of electron loops on photons when the energies of the photons are much smaller than the electron mass. Since the influence action is described in the regime $m^2 \gg R$ by a local expansion, it can be separated into two parts: $S_\mathrm{IF}^\mathrm{ren}[g^+,g^-] = S_\mathrm{eff}[g^+] - S_\mathrm{eff}[g^-]$. $S_\mathrm{eff}[g]$ coincides with the result of integrating out the massive fields in the \emph{in-out} formalism and it is given by
\begin{equation}
\fl
S_\mathrm{eff}[g] = \int d^4x \sqrt{-g} \left[ a\, m^4 + b\, m^2 R
+ c\, R^2 + d\, C^{\mu\nu\rho\sigma} C_{\mu\nu\rho\sigma}
+ O (R^3/m^2) \right]
\label{4.1} ,
\end{equation}
where $a$, $b$, $c$ and $d$ are dimensionless parameters, $C^{\mu\nu\rho\sigma}$ is the Weyl tensor and $O (R^3/m^2)$ denotes terms of order $1/m^2$ or higher, which involve cubic or higher powers of the curvature but also terms with covariant derivatives starting with those quadratic in the curvature and with two covariant derivatives. Note that terms corresponding to a total divergence (such as the $\Box R$ term) or a topological invariant have not been included. That is also the reason why only two of the three possible terms quadratic in the curvature have been considered (the third one being $R_{\mu\nu}R^{\mu\nu}$) since there is a linear combination of the three which corresponds to the Gauss-Bonnet topological invariant.
The first four terms that appear on the right-hand side of (\ref{4.1}) are in general multiplied by divergent coefficients in the bare influence action;%
\footnote{However, the divergent coefficients of the Ricci scalar and a certain linear combination of the $R^2$ and $C^{\mu\nu\rho\sigma} C_{\mu\nu\rho\sigma}$ terms vanish when $\nu = 0$.}
the renormalized influence action is rendered finite by including suitable counterterms in the bare gravitational action, as explained in section~\ref{sec2.3}. (The parameters $a$ and $b$ exhibit a $\mu$ dependence which is compensated by the $\mu$ dependence of $1/G$ and $\Lambda / G$ in the renormalized Einstein-Hilbert action.)
There is, nevertheless, an inherent ambiguity in the finite part of the coupling constants for those counterterms, which can only be determined experimentally. On the other hand, that is not the case for terms of order $1/m^2$ and higher, which do not need to be renormalized and can be explicitly computed using a quasi-local expansion as described above. The value of those coefficients can then be determined exactly assuming a vanishing contribution from the bare gravitational action, and even if there are contributions from physics at higher energy scales, they will be highly suppressed provided that there is a sufficiently large separation of scales.

By functionally differentiating (\ref{4.1}) with respect to the metric one can obtain the expectation value of the stress tensor operator for the massive field, which is clearly conserved since $S_\mathrm{eff}[g]$ is diffeomorphism invariant. The first two terms simply correspond to a finite renormalization of the gravitational coupling constant and the cosmological constant, which is already reabsorbed in their measured values. Furthermore, the terms quadratic in the curvature give no contribution when considering the backreaction of the massive field on the dynamics of small perturbations around de Sitter spacetime. This is because when using the perturbative treatment described in section~(\ref{sec2.4}), one needs to evaluate the stress tensor on the classical de Sitter background, and the functional derivatives of terms quadratic in the curvature, whose results can be found in \cite{birrell94}, vanish when evaluated for a de Sitter metric. Hence, through the order considered in (\ref{4.1}) there is no backreaction effect due to very massive fields in de Sitter. The solutions of the semiclassical equation for spatially isotropic perturbations simply correspond to a de Sitter spacetime associated with a cosmological constant corrected by the finite renormalization due to the first term on the right-hand side of (\ref{4.1}). If this renormalization is already included in the classical value of the cosmological constant, all the perturbed solutions are just time translations of the classical one and, thus, physically equivalent.

Several remarks about the state of the fields are in order. Although there is in general no unambiguous and preferred notion of vacuum and particle in curved spacetime, the quasi-local approximation for a field with a mass sufficiently large compared to the curvature defines a distinguished family of states which correspond to a natural choice of vacuum. For a RW spacetime they coincide with the notion of adiabatic vacua of infinite order that can be introduced in that case, and the number of particles of one state with respect to another state of the family is exponentially suppressed by a factor roughly of order $\exp(-m/H)$. If one starts with one these states and considers its evolution in a spacetime region where the condition $m \gg H$ is satisfied everywhere, it will remain within the same class of vacua and, thus, essentially unexcited. On the other hand, if the evolution involves a region where the condition $m \gg H$ is not satisfied, the state in the region where the condition is satisfied will in general be excited. However, it can still be described in terms of the natural notions of vacuum and particle excitation associated with the adiabatic vacua of infinite order. The backreaction of the field on the spacetime geometry can then be understood as a combination of vacuum polarization effects and the contribution from the energy density of the created particles. While the backreaction due to the vacuum polarization effects is properly taken into account by (\ref{4.1}) and the methods described in this section, the effect of the created particles can be simply described in terms of a non-relativistic perfect fluid (at least for free fields).

The RG running of the cosmological constant (sometimes together with an RG running of the gravitational coupling constant) due to massive fields in the decoupling regime (corresponding to $m^2 \gg H^2$) has been proposed as a natural mechanism leading to a time-dependent dark-energy component \cite{shapiro03,shapiro05}. Those studies relied on general arguments rather than specific models for the quantum fields generating the RG running. On the other hand, the method outlined in this section gives an explicit result for the effective action governing the dynamics of the background spacetime which includes the backreaction effects of a specific microscopic model for the massive quantum fields. As explained above, up to the order usually considered in this kind of studies (the corrections from higher orders are expected to be very small) the only effect is simply a constant renormalization of the gravitational coupling constant (and the cosmological constant), which is already reabsorbed in their measured values. It would be interesting to see what kind of microscopic model for the fields would give rise to the behavior considered in \cite{shapiro03,shapiro05}. While the explicit calculation based on the Schwinger-DeWitt expansion described above was done for free fields (although it could probably be extended to interacting ones), the effective field theory argument based on diffeomorphism symmetry and power counting should apply more generally to interacting theories as well (as far as their gravitational interaction in the decoupling regime is concerned).

\section{Discussion}
\label{sec5}

In this paper we have computed the one-loop vacuum polarization for non-conformal scalar fields in a general spatially-flat RW background. We obtained explicit analytical results for two different regimes corresponding to weakly and strongly non-conformal fields. The \emph{weakly} non-conformal case corresponds to both a small mass with $m^2 \ll H^2$ and a small curvature-coupling parameter $\nu$, which characterizes the deviation from the case of conformal coupling to the curvature. In this case we introduced a perturbative expansion in terms of $m^2 / H^2$ and $\nu$, and obtained an exact expression through quadratic order for the influence action, from which the quantum expectation value of the stress tensor operator for a family of Gaussian states can be derived. This has then been applied to studying the evolution of spatially-isotropic perturbations around de Sitter spacetime when the backreaction due to the non-conformal field is self-consistently included, which corresponds to solving the equations of semiclassical gravity, and we have calculated explicitly the solutions for all times, which are given by (\ref{3.30}) and (\ref{3.42}). There is a self-consistent solution, associated with the Bunch-Davies vacuum for the quantum field, with an effective cosmological constant slightly shifted from its classical value due to the vacuum polarization effects. Furthermore, we have found that this solution is stable under spatially-isotropic perturbations since the perturbed solutions tend to it at late times.%
\footnote{It should be emphasized that a
complete analysis of the backreaction problem in de Sitter spacetime
and its stability should take into account the effect of the quantum
metric fluctuations as well. Including the metric fluctuations is
certainly a crucial aspect, and some steps in that direction are
briefly discussed below. However, given the complexity of a completely
satisfactory treatment involving the quantized metric perturbations,
it is important to make sure that there are no significant effects
even when they are not taken into account, especially because such
effects have actually been suggested by a number of studies in the
literature.}
On the other hand, for \emph{strongly} non-conformal fields with a large mass the influence action for an arbitrary geometry can be expressed as a local expansion in terms of powers of the curvature over the mass squared, and it is independent of the initial state of the field. The lowest order terms simply correspond to a constant renormalization of the cosmological constant and the gravitational coupling constant which is already reabsorbed in their observed values, whereas the contributions from the higher-order terms only give rise to small corrections. In particular, for small spatially-isotropic perturbations around de Sitter the correction due to the terms quadratic in the curvature actually vanishes.
These results for fields with a mass much larger than the inverse of the typical radius of curvature provide a specific example with which dark energy models based on the RG running of the cosmological constant due to massive fields in the decoupling regime can be compared.

When solving the semiclassical Friedmann equation, we introduced a perturbative expansion in powers of $(l_\mathrm{p} H)^2$. Its purpose was to obtain a fairly
accurate description for phenomena involving length-scales much larger
than the Planck length while discarding spurious solutions involving
Planckian scales, where the effective field theory approach that we
have been using breaks down. Furthermore, employing this kind of perturbative treatment made it possible to obtain explicit analytic solutions. However, truncating the perturbative expansion
for the solutions (rather than doing so at the level of the equation
of motion and then solving it exactly) can sometimes miss the right
long-time behavior. In any case, our perturbative calculation would still signal when that is happening: one would find a solution that grows at late times up to a point where the perturbative expansion can no longer be applied. On the other hand, if the perturbative solution remains small at all times, it means that it was safe to use the perturbative treatment.

An important point concerning local terms proportional to the logarithm of the scale factor should be emphasized. As seen in section~\ref{sec3}, a local term proportional to the logarithm of the scale factor appears generically in the influence action [see equation (\ref{3.6})]. For an expanding spacetime, and especially for an exponentially expanding one like de Sitter, the contribution from this term can dominate at sufficiently late times over all the other local terms. In particular, it gives rise to a logarithmically growing term in the expression for the expectation value of the stress tensor which would prevent the existence of a self-consistent de Sitter solution and would imply
an increasing deviation from the classical de Sitter background. However, it turns out that this growing contribution is cancelled out by a similar contribution form the non-local term. This can be explicitly seen in our result for the stress tensor in the example of a massive field with conformal coupling. In that case the corresponding local term in the action is proportional to $m^4$ and has the same form as a cosmological constant term times the additional $\ln a$ factor.
The logarithmic term that appears explicitly in (\ref{3.16}) is cancelled out by the first term on the right-hand side of (\ref{3.20}), which contains the contributions from the non-local term. As a consequence, the final expression for the stress tensor, given by (\ref{3.22}), exhibits no logarithmically growing term (there can be other logarithmic terms, but relatively suppressed by negative powers of the scale factor). Furthermore, we find that there is a self-consistent de Sitter solution of the backreaction equation and it is stable under spatially isotropic perturbations.
The situation is similar in the other cases. For a massless field with non-conformal coupling there is a local term with the logarithm of the scale factor multiplying the square of the Ricci scalar.%
\footnote{Note that this term gives no logarithmically growing contribution to the stress tensor evaluated on a de Sitter background for the same reason why the local $R^2$ term gives a vanishing contribution to the stress tensor when evaluated on a de Sitter background. Nevertheless, it will give in general a logarithmically growing contribution when considering a background different from de Sitter (more specifically, whenever a background with $\Box R \neq 0$ is considered), or when considering higher orders in $(l_\mathrm{p} H)^2$.}
Finally, for a massive field with non-conformal coupling one has the two terms of the previous two cases plus a term proportional to $m^2$ with $\ln a$ multiplying the Ricci scalar.
In all those cases there is a cancellation of the logarithmically growing contribution from the local terms proportional to $\ln a$ and a similar contribution from the non-local term. In fact, this phenomenon is fairly general: one can see that something analogous also happens when considering spatially anisotropic and inhomogeneous perturbations \cite{campos94,campos96}.
In light of these results, it may be worth revisiting related studies where the non-local terms where not considered. In particular, a local approximation for the effective action which excludes the non-local terms (but includes the local terms proportional to $\ln a$)
is commonly employed in the literature and often leads to significant deviations from the de Sitter solution due to the backreaction of massive fields treated perturbatively.
Our explicit calculation of the effective action (including the non-local terms) can contribute to a better understanding of the validity of the local approximation in those cases.

In this paper we have calculated the effective action and solved the backreaction equations for free scalar fields. However, it is fairly straightforward to extend our calculations to fermions or vector fields. We expect the results to be similar to the scalar case, and the main conclusions to remain the same. Furthermore, it should be noted that our results can also be useful to compare with similar analysis of the backreaction due to graviton one-loop effects. Although we did not quantize the metric (and did not have to deal with the ambiguities associated with the gauge-fixing of the quantum metric perturbations), our results for the effective action have mathematically the same form as that in some studies of the graviton case. In particular, our influence action for the massless non-conformally coupled case has the same form as the effective action in \cite{espriu05}. Similarly, our influence action for the weakly massive and conformally coupled case has the same form as the dominant contribution considered in \cite{cabrer07} if one replaces the cosmological constant in their result with our $m^2$ (their full expression including subdominant terms in the infrared limit would be analogous to our weakly massive and non-conformally coupled case). Note that our result was derived for $m^2 \ll H^2$ whereas their cosmological constant $\Lambda$ is of the same order as $H^2$. Nevertheless, this condition does not play any role either in the derivation of the stress tensor from the effective action or when solving the backreaction equation.

We have studied the backreaction of the quantum fields on the
dynamics of the spacetime geometry within the framework of
semiclassical gravity, which can be understood as a mean field
approximation where the mean gravitational field is described by a
classical metric whereas its quantum fluctuations are not considered.
In order to study the quantum fluctuations of the gravitational field
one can consider the metric perturbations around a background geometry
corresponding to the semiclassical gravity solution and quantize them
within a low-energy effective field theory approach to quantum gravity
\cite{donoghue94a,donoghue94b,donoghue97,burgess04}. So far this
approach has been mostly applied to weak-field problems
\cite{bjerrum03}, but it seems particularly interesting to
extend its application to strong-field situations involving black
holes \cite{hu07a,roura07a,hu07b} and cosmological spacetimes \cite{weinberg05,weinberg06}. The
stochastic gravity formalism \cite{calzetta94,martin99a,martin99b,hu03a,hu04a} can be a useful tool
in this respect since one can prove its equivalence to a quantum
treatment of the metric perturbations if graviton loops are neglected,
which can be formally justified in a large $N$ expansion for a large
number of matter fields \cite{hu04b}. A central object in this
formalism is the symmetrized connected two-point function of the
stress tensor operator for the quantum matter fields, which determines
the metric fluctuations induced by the quantum fluctuations of the
matter fields. Such an object has been computed for a massless
minimally coupled field evolving in a de Sitter background spacetime
and the fluctuations of the stress tensor were found to be comparable
to its expectation value \cite{roura99b}. Therefore, studying in
detail the quantum fluctuations of the metric in this context
constitutes a natural extension of our work worth pursuing. The
results obtained here would still be relevant in that case because
they provide the right background around which the metric should be
perturbed and quantized.

We close this section with a brief discussion of the relationship
between our results and the linearization instability for metric
perturbations around de Sitter spacetime coupled to a scalar field
found in \cite{losic06}, where it was concluded that it is only
consistent to consider de Sitter invariant states for the quantum
field.  This conclusion does not directly affect our analysis because
we did not consider fluctuations of the metric and studied only the
dynamics of the mean geometry, which couples to the expectation value
of the stress tensor operator of the matter field. The expectation
value of the stress tensor for the class of states that we have
considered in this paper, which are spatially homogenous and
isotropic, automatically satisfies the linearization stability
constraint given by equation (44) in \cite{losic06}. It is when
considering the quantum fluctuations of the metric that the
linearization stability condition imposes additional restrictions on
the state of the matter field because in that case the condition must be imposed on the $n$-point correlation
functions of the stress tensor as well,
and this implies that the state of the field must be de Sitter
invariant.%

\ack
We are grateful to Paul Anderson, Daniel Arteaga, Diego Blas, Dom\`enec Espriu, Jaume Garriga, Carmen Molina-Par\'{\i}s, Emil Mottola, Ilya Shapiro and Joan Sol\`a for interesting discussions. This work has been partly supported by the Research Projects MEC FPA2007-66665C02-02 and DURSI 2005-SGR-00082. A.~R.\ is supported by LDRD funds from Los Alamos National Laboratory.

\appendix

\section*{Appendix}
\label{app}

\setcounter{section}{1}

As has been pointed out in section \ref{sec3}, the renormalized expectation value of the trace of the stress tensor may suffer from initial time divergences unless we choose an appropriate state for the quantum field. In this appendix we will carefully examine this point. For simplicity, we will restrict ourselves to the massless case, $m=0$, $\nu\ne 0$. However, no assumptions on the size of the $\nu$ parameter will be made here, so the following results apply for both the weakly and the strongly non-conformal regimes.
Throughout the appendix we will follow closely the definitions and results for fourth-order adiabatic vacua of \cite{habib00}.

We will make use of an alternative expression for the renormalized expectation value of the trace of the stress tensor. Consider the mode functions $f_k$, solution of (\ref{2.6}) with \emph{any} initial conditions consistent with the Wronskian condition (\ref{2.6b}), not necessarily of the form (\ref{2.8}).  The bare expectation value of the trace of the stress tensor in the vacuum state associated with this set of modes can be written as (see, for instance, \cite{lindig99})
\begin{equation}
\langle T^{\mu}_\mu\rangle=-\frac{6\nu}{2\pi^2a^4}\int_{0}^\infty d k\, k^2
\left[\dot h|f_k|^2+ 2 h\, |f_k| \frac{d |f_k|}{d\eta} -|\dot f_k |^2
+\Omega_k^2|f_k|^2\right],
\label{a1}
\end{equation}
where we have used the notations $\Omega_k^2\equiv k^2+M^2$ and $h\equiv\dot a/a$. In order to renormalize this expectation value, it is convenient to introduce the WKB mode functions. The most general mode function satisfying the Wronskian condition (\ref{2.6b}) can be written as
\begin{equation}
f_k(\eta)=\frac{1}{\sqrt{2W_k}}\exp\left[-i\int^{\eta}_{\eta_i}d\eta'\,W_k(\eta')\right],
\label{a2}
\end{equation}
with real and positive time-dependent frequency $W_k$. [We are assuming that $f_k$ is real and positive at the initial time, which implies no loss of generality because of global phase freedom.] With this change of variables, the mode equation (\ref{2.6}) takes the form
\begin{equation}
W_{k}^2=\Omega_k^2-\frac{1}{2}\left[\frac{\ddot W_k}{W_k}-\frac{3}{2}\left(\frac{\dot W_k}{W_k}\right)^2\right].
\label{a3}
\end{equation}
Solving this equation iteratively, starting with $W_k=\Omega_k$, one obtains an adiabatic expansion for $W_k$. This is the WKB solution of (\ref{a3}), and the mode associated with it through (\ref{a2}) is a WKB mode. The WKB frequency admits the following expansion in inverse powers of $k$:
\begin{equation}
W_{k}=k+\frac{M^2}{2k}-\frac{M^4+ (d^2M^2 / d\eta^2)}{8k^3}+O(k^{-5}).
\label{a4}
\end{equation}
The adiabatic approximation of fourth order, which contains up to fourth-order derivatives of the scale factor, is obtained by truncating the iterative procedure after the second iteration,
\begin{equation}
\left({W_{k}^{(4)}}\right)^2=\Omega_k^2-\frac{1}{2}
\left[\frac{\ddot \Omega_k}{\Omega_k}
-\frac{3}{2}\left(\frac{\dot \Omega_k}{\Omega_k}\right)^2\right],
\label{a5}
\end{equation}
and differs from the exact solution $W_k(\eta)$ only in $O(k^{-5})$ terms. 
The bare expectation value (\ref{a1}) can now be renormalized by the so-called adiabatic subtraction procedure \cite{fulling74,parker74,bunch80}, which consists of subtracting the same expectation value computed with the fourth-order adiabatic approximation to the WKB modes:
\begin{equation}
\langle T^{\mu}_\mu\rangle_{\rm ren}=\langle T^{\mu}_\mu\rangle
-\langle T^{\mu}_\mu\rangle^{(4)},
\label{a7}
\end{equation}
where $\langle T^{\mu}_\mu\rangle^{(4)}$ is the trace (\ref{a1}) associated with the modes (\ref{a2}), with time-dependent frequency $W_k$ given by (\ref{a5}). This procedure has been shown to be equivalent to
covariant renormalization methods specialized to
RW spacetimes \cite{birrell78,anderson87}.

\subsection*{Physical vacua}

What kind of initial conditions should the modes $\{f_k\}$ satisfy in order for the renormalized expectation value of the trace $\langle T^{\mu}_\mu\rangle_{\rm ren}$ to be finite at the initial time $\eta_{\rm i}$? According to (\ref{a2}), the most general initial conditions that fulfill the Wronskian condition (\ref{2.6b}) are
\begin{eqnarray}
f_k(\eta_{\rm i})&=&\frac{1}{\sqrt{2w_k}}\nonumber\\
\dot f_k(\eta_{\rm i})&=&\left(-i w_k+\frac{v_k}{2}\right)f_k(\eta_i),
\label{a8}
\end{eqnarray}
where $w_k$ and $v_k$ are two arbitrary real functions of $k$ (with $w_k>0$). Substituting these equalities into (\ref{a1}) and (\ref{a7}), we get that the renormalized expectation value of the trace at the initial time is
\begin{eqnarray}
\fl
\langle T^{\mu}_\mu\rangle_{\rm ren}(\eta_{\rm i})
= -\frac{3\nu}{2\pi^2a^{4}_{\rm i}}\int_{0}^\infty d k\, k^2
&& \left\{\dot h_{\rm i}\left[\frac{1}{w_{k}}
-\frac{1}{W_{k{\rm i}}^{(4)}}\right]
+h_{\rm i}\left[\frac{v_{k}}{w_{k}}+\frac{\dot W_{k{\rm i}}^{(4)}}
{\left(W_{k{\rm i}}^{(4)}\right)^2}\right]
-\left[w_k-W_{k{\rm i}}^{(4)}\right]\right.
\nonumber\\
&&\left. +\, \Omega_{k{\rm i}}^2\left[\frac{1}{w_{k}}
-\frac{1}{W_{k{\rm i}}^{(4)}}\right]-\left[\frac{v_{k}^2}{4w_{k}}
-\frac{\left(\dot W_{k{\rm i}}^{(4)}\right)^2}
{4\left(W_{k{\rm i}}^{(4)}\right)^3}\right]\right\},
\label{a9}
\end{eqnarray}
where the subscript ``i'' indicates that the function is evaluated at the initial time $\eta_{\rm i}$. This integral is finite provided that the integrand falls off faster than $k^{-1}$ as $k\to\infty$, which is achieved if
\begin{eqnarray}
\left|w_k-W_{k{\rm i}}\right|&<&O(k^{-3}) \label{b14a}\nonumber \\
\left|v_k+\frac{\dot W_{k{\rm i}}}{W_{k{\rm i}}}\right|&<&O(k^{-2})
\label{a10}
\end{eqnarray}
when $k\to\infty$. Here we have used the fact that $W_k$ and $W_{k}^{(4)}$ differ only by $O(k^{-5})$ terms. These are the constraints that select, among all the possible homogeneous and isotropic vacua, the physical ones, namely the ones that keep the trace finite at the initial time. They correspond to the so-called fourth-order adiabatic vacua.

\subsection*{Our vacua}

Let us now see how these constraints affect our particular class of vacuum states. In other words: among our vacua, which are the physical ones? Our vacuum states are associated with the initial conditions defined by (\ref{2.7})-(\ref{2.10}). Note that, in fact, in the limit $k\to\infty$ the modes $\{f_{k}^\Psi\}$ are just the WKB modes of equation (\ref{2.9}),
\begin{equation}
f_{k}^\Psi(\eta)=\frac{1}{\sqrt{2W_{k}^\Psi}}\exp\left[-i\int^{\eta}_{\eta_i}d\eta'\,W_{k}^\Psi(\eta')\right],
\label{a11}
\end{equation}
where $W_{k}^\Psi$ is given by (\ref{a4}) with $M_{\Psi}$ replacing $M$.  The reason for this is that, for $k\to\infty$, both sides of the equation above satisfy the mode equation (\ref{2.9}), and both sides reduce to standard plane waves when $\eta\to-\infty$. Therefore, our particular initial conditions can be written in the form (\ref{a8}), with $w_k$ and $v_k$ given in this case by
\begin{eqnarray}
w_k=W_{k{\rm i}}^\Psi \\
v_k=-\frac{\dot{W}_{k{\rm i}}^\Psi}{W_{k{\rm i}}^\Psi}
\label{a12}
\end{eqnarray}
for $k\to\infty$. The constraints (\ref{a10}), together with the expansion (\ref{a4}), then imply
\begin{eqnarray}
\fl
\left|\frac{1}{2k}[M_{\Psi{\rm i}}^2-M_{\rm i}^2]
-\frac{1}{8k^3}\Big[M_{\Psi{\rm i}}^4-M_{\rm i}^4
+\left(d^2 M_{\Psi}^2 / d\eta^2\right)_{\rm i}
-\left(d^2 M^2 / d\eta^2\right)_{\rm i} \Big]+O(k^{-5})\right|<O(k^{-3})\nonumber \\
\fl
\left|\frac{1}{2k^2}\Big[\left(d\, M_{\Psi}^2 / d\eta\right)_{\rm i}
- \left(d\, M^2 / d\eta\right)_{\rm i}\Big]+O(k^{-4})\right|<O(k^{-2}). \nonumber
\label{a13}
\end{eqnarray}
In order for these inequalities to be satisfied, we must impose
\begin{equation}
\fl
M_{\Psi{\rm i}}^2=M_{\rm i}^2\qquad \left(d\, M_{\Psi}^2 / d\eta\right)_{\rm i}
=\left(d\, M^2 / d\eta\right)_{\rm i}
\qquad \left(d^2 M_{\Psi}^2 / d\eta^2\right)_{\rm i}
=\left(d^2 M^2 / d\eta^2\right)_{\rm i}.
\label{a14}
\end{equation} 
which imply precisely the conditions in (\ref{3.33}), which were obtained by other means in section~\ref{sec3}. Therefore, the requirement that the scale factors $a_\Psi$ and $a$ are matched smoothly enough selects, among all the states considered in this paper, the fourth-order adiabatic vacua.

We close this appendix by mentioning an alternative method of calculating the influence action (but entirely equivalent to that of section~\ref{sec3}) provided in \cite{roura99a}. The approach, which is based on decomposing the
field in spatial Fourier modes, computing the unitary evolution
operator for each mode perturbatively in the interaction picture, and
summing over all the modes at the end, can be useful when considering
more general initial states at a finite initial time $\eta_\mathrm{i}$
which are not necessarily of the form described in section~\ref{sec2.1}.

\section*{References}


\begin{thebibliography}{10}
\providecommand{\url}[1]{\texttt{#1}}
\providecommand{\urlprefix}{URL }
\providecommand{\eprint}[2][]{\url{#2}}

\bibitem{guth81}
Guth A~H 1981 \textit{Phys. Rev. D} \textbf{23} 347

\bibitem{linde82a}
Linde A~D 1982 \textit{Phys. Lett. B} \textbf{108} 389

\bibitem{albrecht82}
Albrecht A, Steinhardt P~J 1982 \textit{Phys. Rev. Lett.} \textbf{48} 1220

\bibitem{linde83a}
Linde A~D 1983 \textit{Phys. Lett. B} \textbf{129} 177

\bibitem{linde90}
Linde A~D 1990 \textit{Particle physics and inflationary cosmology} Harwood
  Academic, Amsterdam

\bibitem{smoot92}
Smoot G~F, \textit{et~al.} 1992 \textit{Astrophys. J.} \textbf{396} L1

\bibitem{bennett03}
Bennett C~L, \textit{et~al.} 2003 \textit{Astrophys. J. Suppl.} \textbf{148} 1

\bibitem{peiris03}
Peiris H~V, \textit{et~al.} 2003 \textit{Astrophys. J. Suppl.} \textbf{148} 213

\bibitem{spergel07}
Spergel D~N, \textit{et~al.} 2007 \textit{Astrophys. J. Suppl.} \textbf{170}
  377

\bibitem{perlmutter99}
Perlmutter S, \textit{et~al.} 1999 \textit{Astrophys. J.} \textbf{517} 565

\bibitem{riess98}
Riess A~G, \textit{et~al.} 1998 \textit{Astrophys. J.} \textbf{116} 1009

\bibitem{seljak06}
Seljak U, Slosar A, McDonald P 2006 \textit{JCAP} \textbf{0610} 014

\bibitem{tegmark06}
Tegmark M, \textit{et~al.} 2006 \textit{Phys. Rev. D} \textbf{74} 123507

\bibitem{giannantonio06}
Giannantonio T, \textit{et~al.} 2006 \textit{Phys. Rev. D} \textbf{74} 063520

\bibitem{eisenstein05}
Eisenstein D~J 2005 \textit{Astrophys. J.} \textbf{633} 560

\bibitem{tsamis96a}
Tsamis N~C, Woodard R~P 1996 \textit{Nucl. Phys. B} \textbf{474} 235

\bibitem{tsamis97}
Tsamis N~C, Woodard R~P 1997 \textit{Ann. Phys. (NY)} \textbf{253} 1

\bibitem{mukhanov97}
Mukhanov V~F, Abramo L~R~W, Brandenberger R~H 1997 \textit{Phys. Rev. Lett.}
  \textbf{78} 1624

\bibitem{abramo97}
Abramo L~R~W, Brandenberger R~H, Mukhanov V~F 1997 \textit{Phys. Rev. D}
  \textbf{56} 3248

\bibitem{abramo99}
Abramo L~R~W, Woodard R~P 1999 \textit{Phys. Rev. D} \textbf{60} 044010

\bibitem{losic05}
Losic B, Unruh W~G 2005 \textit{Phys. Rev. D} \textbf{72} 123510

\bibitem{giddings05}
Giddings S~B, Marolf D, Hartle J~B 2006 \textit{Phys. Rev. D} \textbf{74}
  064018

\bibitem{abramo02b}
Abramo L~R~W, Woodard R~P 2002 \textit{Phys. Rev. D} \textbf{65} 063515

\bibitem{geshnizjani02}
Geshnizjani G, Brandenberger R 2002 \textit{Phys. Rev. D} \textbf{66} 123507

\bibitem{garriga08}
Garriga J, Tanaka T 2008 \textit{Phys. Rev. D} \textbf{77} 024021

\bibitem{tsamis07}
Tsamis N~C, Woodard R~P 2007 Reply to `{C}an infrared gravitons screen
  {$\Lambda$}?' \eprint{arXiv:0708.2004v2 [hep-th]}

\bibitem{fischetti79}
Fischetti M~V, Hartle J~B, Hu B~L 1979 \textit{Phys. Rev. D} \textbf{20} 1757

\bibitem{starobinsky80}
Starobinsky A~A 1980 \textit{Phys. Lett. B} \textbf{91} 99

\bibitem{vilenkin85}
Vilenkin A 1985 \textit{Phys. Rev. D} \textbf{32} 2511

\bibitem{simon91}
Simon J~Z 1991 \textit{Phys. Rev. D} \textbf{43} 3308

\bibitem{parker93}
Parker L, Simon J~Z 1993 \textit{Phys. Rev. D} \textbf{47} 1339

\bibitem{flanagan96}
Flanagan E~E, Wald R~M 1996 \textit{Phys. Rev. D} \textbf{54} 6233

\bibitem{burgess04}
Burgess C~P 2004 \textit{Living Rev. Rel.} \textbf{7} 5

\bibitem{simon92}
Simon J~Z 1992 \textit{Phys. Rev. D} \textbf{45} 1953

\bibitem{hawking01}
Hawking S~W, Hertog T, Reall H~S 2001 \textit{Phys. Rev. D} \textbf{63} 083504

\bibitem{shapiro02}
Shapiro I~L, Sol\`a J 2002 \textit{Phys. Lett. B} \textbf{530} 10

\bibitem{pelinson03a}
Pelinson A, Shapiro I, Takakura F 2003 \textit{Nucl. Phys. B} \textbf{648} 417

\bibitem{espriu05}
Espriu D, Multam\"aki T, Vagenas E~C 2005 \textit{Phys. Lett. B} \textbf{628}
  197

\bibitem{cabrer07}
Cabrer J~A, Espriu D 2007 Secular effects on inflation from one-loop quantum
  gravity \eprint{arXiv:0710.0855 [gr-qc]}

\bibitem{birrell94}
Birrell N~D, Davies P~C~W 1994 \textit{Quantum fields in curved space}
  Cambridge University Press, Cambridge

\bibitem{wald94}
Wald R~M 1994 \textit{Quantum field theory in curved spacetime and black hole
  thermodynamics} The University of Chicago Press, Chicago

\bibitem{bunch78a}
Bunch T~S, Davies P~C~W 1978 \textit{Proc. R. Soc. London A} \textbf{360} 117

\bibitem{dowker76}
Dowker J~S, Critchley R 1976 \textit{Phys. Rev. D} \textbf{13} 3224

\bibitem{wada83}
Wada S, Azuma T 1983 \textit{Phys. Lett. B} \textbf{132} 313

\bibitem{anderson00}
Anderson P~R, Eaker W, Habib S, Molina-Par\'{\i}s C, Mottola E 2000
  \textit{Phys. Rev. D} \textbf{62} 124019

\bibitem{habib00}
Habib S, Molina-Par\'{\i}s C, Mottola E 2000 \textit{Phys. Rev. D} \textbf{61}
  024010

\bibitem{isaacson91}
Isaacson J~A, Rogers B 1991 \textit{Nucl. Phys. B} \textbf{364} 381

\bibitem{rogers92}
Rogers B, Isaacson J~A 1992 \textit{Nucl. Phys. B} \textbf{368} 415

\bibitem{busch92}
Busch C 1992 The semiclassical stability of de {S}itter space-time
  \eprint{DESY-92-131, ITP-UH-9-92; arXiv:0803.3204 [gr-qc]}

\bibitem{perez-nadal07}
P\'erez-Nadal G, Roura A, Verdaguer E 2007 Stability of de {S}itter spacetime
  under isotropic perturbations in semiclassical gravity
  \eprint{arXiv:0712.2282 [gr-qc]}

\bibitem{misner73}
Misner C~W, Thorne K~S, Wheeler J~A 1973 \textit{Gravitation} Freeman, San
  Francisco

\bibitem{martin99b}
Mart\'{\i }n R, Verdaguer E 1999 \textit{Phys. Rev. D} \textbf{60} 084008

\bibitem{hu04a}
Hu B~L, Verdaguer E 2004 \textit{Living Rev. Rel.} \textbf{7} 3

\bibitem{donoghue97}
Donoghue J~F 1999 T~Piran, R~Ruffini (eds.) \textit{The Eighth Marcel Grossmann
  Meeting} World Scientific, Singapore \eprint{gr-qc/9712070}

\bibitem{hu07b}
Hu B~L, Roura A 2007 \textit{Phys. Rev. D} \textbf{76} 124018

\bibitem{campos94}
Campos A, Verdaguer E 1994 \textit{Phys. Rev. D} \textbf{49} 1861

\bibitem{campos97}
Campos A, Verdaguer E 1997 \textit{Int. J. Theor. Phys.} \textbf{36} 2525

\bibitem{calzetta97c}
Calzetta E, Campos A, Verdaguer E 1997 \textit{Phys. Rev. D} \textbf{56} 2163

\bibitem{leibbrandt75}
Leibbrandt G 1975 \textit{Rev. Mod. Phys.} \textbf{47} 849

\bibitem{schwartz66}
Schwartz L 1966 \textit{Th\'eory des distributions} Hermann, Paris

\bibitem{frolov84}
Frolov V~P, Zelnikov A~I 1984 \textit{Phys. Rev. D} \textbf{29} 1057

\bibitem{itzykson80}
Itzykson C, Zuber J~B 1980 \textit{Quantum field theory} McGraw-Hill, New York

\bibitem{shapiro03}
Shapiro I~L, Sol\`a J, {Espa\~na-Bonet} C, Ruiz-Lapuente P 2003 \textit{Phys.
  Lett. B} \textbf{574} 149

\bibitem{shapiro05}
Shapiro I~L, Sol\`a J, Stefancic H 2005 \textit{JCAP} \textbf{0501} 012

\bibitem{campos96}
Campos A, Verdaguer E 1996 \textit{Phys. Rev. D} \textbf{53} 1927

\bibitem{donoghue94a}
Donoghue J~F 1994 \textit{Phys. Rev. Lett.} \textbf{72} 2996

\bibitem{donoghue94b}
Donoghue J~F 1994 \textit{Phys. Rev. D} \textbf{50} 3874

\bibitem{bjerrum03}
Bjerrum-Bohr N~E~J, Donoghue J~F, Holstein B~R 2003 \textit{Phys. Rev. D}
  \textbf{68} 084005 erratum-ibid. D {\bf 71}, 069904 (2005)

\bibitem{hu07a}
Hu B~L, Roura A 2007 \textit{Int. J. Theor. Phys.} \textbf{46} 2204

\bibitem{roura07a}
Roura A 2007 \textit{J. Phys. A} \textbf{40} 7075

\bibitem{weinberg05}
Weinberg S 2005 \textit{Phys. Rev. D} \textbf{72} 043514

\bibitem{weinberg06}
Weinberg S 2006 \textit{Phys. Rev. D} \textbf{74} 023508

\bibitem{calzetta94}
Calzetta E, Hu B~L 1994 \textit{Phys. Rev. D} \textbf{49} 6636

\bibitem{martin99a}
Mart\'{\i }n R, Verdaguer E 1999 \textit{Phys. Lett. B} \textbf{465} 113

\bibitem{hu03a}
Hu B~L, Verdaguer E 2003 \textit{Class. Quant. Grav.} \textbf{20} R1

\bibitem{hu04b}
Hu B~L, Roura A, Verdaguer E 2004 \textit{Phys. Rev. D} \textbf{70} 044002

\bibitem{roura99b}
Roura A, Verdaguer E 1999 \textit{Int. J. Theor. Phys.} \textbf{38} 3123

\bibitem{losic06}
Losic B, Unruh W~G 2006 \textit{Phys. Rev. D} \textbf{74} 023511

\bibitem{lindig99}
Lindig J 1999 \textit{Phys. Rev. D} \textbf{59} 064011

\bibitem{fulling74}
Fulling S~A, Parker L 1974 \textit{Ann. Phys. (NY)} \textbf{87} 176

\bibitem{parker74}
Parker L, Fulling S~A 1974 \textit{Phys. Rev. D} \textbf{9} 341

\bibitem{bunch80}
Bunch T~S 1980 \textit{J. Phys. A} \textbf{13} 1297

\bibitem{birrell78}
Birrell N~D 1978 \textit{Proc. R. Soc. Lond. B} \textbf{361} 513

\bibitem{anderson87}
Anderson P~R, Parker L 1987 \textit{Phys. Rev. D} \textbf{36} 2963

\bibitem{roura99a}
Roura A, Verdaguer E 1999 \textit{Phys. Rev. D} \textbf{60} 107503

\end{thebibliography}

\end{document}